\documentclass[journal]{IEEEtran}
\usepackage{amsmath,amsfonts}
\usepackage{algorithmic}
\usepackage{algorithm}
\usepackage{array}
\usepackage[caption=false,font=normalsize,labelfont=sf,textfont=sf]{subfig}
\usepackage{textcomp}
\usepackage{stfloats}
\usepackage{url}
\usepackage{verbatim}
\usepackage{graphicx}
\usepackage{cite}
\begin{document}

\title{Performance Assessment of Hybrid and Digital Irregular Array Configurations for Beyond 100\--GHz Multi\--User MIMO Systems}

\author{Yi\u{g}it Ertu\u{g}rul, Kamil Yavuz Kapusuz, Claude Desset, Sofie Pollin
}



\maketitle

\begin{abstract}
The performance of irregular phased array architectures is assessed in the context of multi-user multiple-input multiple-output~(MU-MIMO) communications operating beyond 100~GHz. Realizing half\--wavelength spaced planar phased arrays is challenging due to wavelength\--integrated circuit~(IC) size conflict at those frequencies where the antenna dimensions are comparable to IC size. Therefore, irregular array architectures such as thinned and clustered arrays are developed to mitigate the wavelength\--IC size conflict. In the thinned arrays, radiating elements are permanently deactivated, while in clustered arrays, neighboring elements are grouped into subarrays. Furthermore, irregular arrays are integrated with hybrid beamforming architectures to manage the complexity introduced by full digital beamforming where a single radio frequency chain is connected per power amplifier. An optimization problem is formulated to determine the optimal arrangement of antenna elements where the trade-off between spectral efficiency~(SE) and sidelobe levels~(SLL) can be tuned. Clustered array configurations are optimized by genetic algorithm and Algorithm\--X based methodologies where the former relies on a randomized search and the latter exploits brute-force search, respectively. Furthermore, a prototype array is designed on a printed circuit board~(PCB) to verify the proposed methodologies through full\--wave simulations. The developed irregular array assessment framework is validated through numerical simulations considering an 8$\times$10 transmit array serving two receivers. To have a fair comparison, clustered arrays with a grouping of two and four elements are compared with thinned arrays with half and quarter thinning ratio, respectively. Thinned array configurations are achieving a peak SE of 3.4~b/s/Hz and 2.13~b/s/Hz while clustered array configurations are achieving a peak SE of 4.53~b/s/Hz and 3.67~b/s/Hz, respectively. The combination of hybrid and irregular array architectures leads to minimal or no performance degradation in the case of hybrid fully connected architectures but severe SE and SLL degradation in the case of hybrid partially connected architectures, respectively.
\end{abstract}


\section{Introduction}
\IEEEPARstart{T}{he} exponential growth in mobile data traffic and the increasing demand for ultra-reliable, low-latency communication are driving the evolution of wireless networks towards the millimeter\--wave (mm-wave) frequency spectrum, particularly beyond 100~GHz~\cite{6gvision,Jornet2024}. This frequency range is considered a promising frontier for the development of future wireless communication systems, such as the upcoming sixth-generation (6G) networks. Leveraging these high frequencies offers the potential for massive bandwidth and data rates that far exceed current capabilities, enabling advanced applications such as immersive augmented reality (AR), virtual reality (VR), holographic communication, and massive machine-type communications (mMTC) for the Internet of Things (IoT). However, operating in the mm-wave spectrum introduces significant challenges, including severe signal attenuation, limited scattering, and increased susceptibility to environmental obstacles. To address these issues, innovative technologies like multi-user multiple-input multiple-output (MU-MIMO) systems are being explored as key enablers to enhance system capacity, improve spectral efficiency (SE), and ensure robust communication in these challenging environments~\cite{5G,802ad}.

At frequencies beyond 100~GHz, the implementation of conventional half-wavelength spaced phased arrays presents significant challenges due to packaging constraints~\cite{Sadhu2019}, as integrated circuit (IC) footprint does not scale with frequency at the same rate as antenna size. This results in a size conflict between the wavelength and IC dimensions in phased arrays operating at these frequencies. To address this, existing array architectures~\cite{Karakuzulu2021,Zhang2024_2} often incorporate subarrays, where a single power amplifier (PA) and a phase shifter drive multiple antennas, or they relax the half-wavelength spacing requirement~\cite{Ahmed2024}. Consequently, irregular array architectures, such as thinned and clustered arrays~\cite{rocca2016unconventional}, have become attractive options to mitigate the wavelength-circuit size conflict.

On the other hand, hybrid and digital beamforming~\cite{Ahmed2018}, where a radio frequency (RF) chain is connected to a plurality of PAs, are critical techniques to overcome the challenges posed by high path loss, limited scattering, and the need for precise beam steering. A combination of irregular arrays and hybrid architectures will pave the way to the realization of such large antenna arrays for future communication systems.

Thinned and clustered array configurations have been optimized to maximize the channel gain or to reduce the sidelobe levels (SLLs) in the context of single\--user MIMO~\cite{Ma2022}. However, the array performance is analyzed separately for SE and SLL. On the other hand, irregular array configurations, which deviate from traditional uniform grid-based antenna arrangements, offer potential advantages~\cite{Anselmi2022} in flexibility, spatial diversity, and array compactness. However, their performance in the context of high-frequency MU-MIMO systems has not been extensively explored.


This article is based on and extends our previous works~\cite{Ertugrul2024, Ertugrul2024_2}, where irregular array architectures were compared considering a MU-MIMO setup in sub\--THz band. The following are the novel contributions of this work,
\begin{itemize}
\item The development of a unified system model to assess the performance of irregular array architectures.
    \item The performance assessment of irregular arrays in combination with hybrid architectures in the context of MU-MIMO.
    \item A genetic algorithm (GA) search tailored to find optimal tetromino\--tiled phased arrays.
    \item The performance validation of the optimized array configurations by full\--wave simulations.
\end{itemize}

The article is organized as follows. Section~\ref{sec:Methods} describes the methods used to assess the performance of hybrid and digital irregular array configurations. Section~\ref{sec:Sysmodel} presents the system model, introduces two key performance metrics, the associated signal model for hybrid architectures, and explains irregular array architectures along with the objective function for the performance assessment. Irregular array configuration generation schemes are presented in Section~\ref{sec:ConfGen}, and the implementation aspects of irregular arrays are discussed in Section~\ref{sec:CST}. Section~\ref{sec:Results} validates the developed methodology with numerical examples, followed by conclusion in Section~\ref{sec:conc}.

\textit{Notations:} Vectors and matrices are denoted by bold lowercase and uppercase letters such as $\mathbf{a}$ and $\mathbf{A}$, respectively. Matrix hermitian and transpose are denoted by $(.)^H$ and $(.)^T$, respectively. Sets are denoted by Calligraphic letters such as $\mathcal{A}$. $\mathbb{C}$ is the set of complex numbers and $\mathbb{E}[.]$ is the expectation operator. Frobenius norm is denoted by $||.||_F$.
\section{Methods/experimental}\label{sec:Methods}
This article focuses on comparing the performance of irregular array architectures in the context of MU-MIMO systems. The design and analysis are conducted with a sub-THz channel model, alongside the relevant antenna array technologies.

The performance comparison is based on Monte Carlo simulations, generating random channel realizations. Various array configurations, including thinned and clustered arrays, were generated through random sampling techniques. To optimize these configurations, a combination of brute-force search, such as Algorithm\--X~\cite{knuth2000} based methods, and randomized search, such as GA-based methods, was employed, respectively. The optimized array designs were then validated using full-wave electromagnetic (EM) simulations to confirm their effectiveness.
\section{System Model}\label{sec:Sysmodel}
\begin{figure*}[t!]
    \centering
    \includegraphics[width=0.245\linewidth]{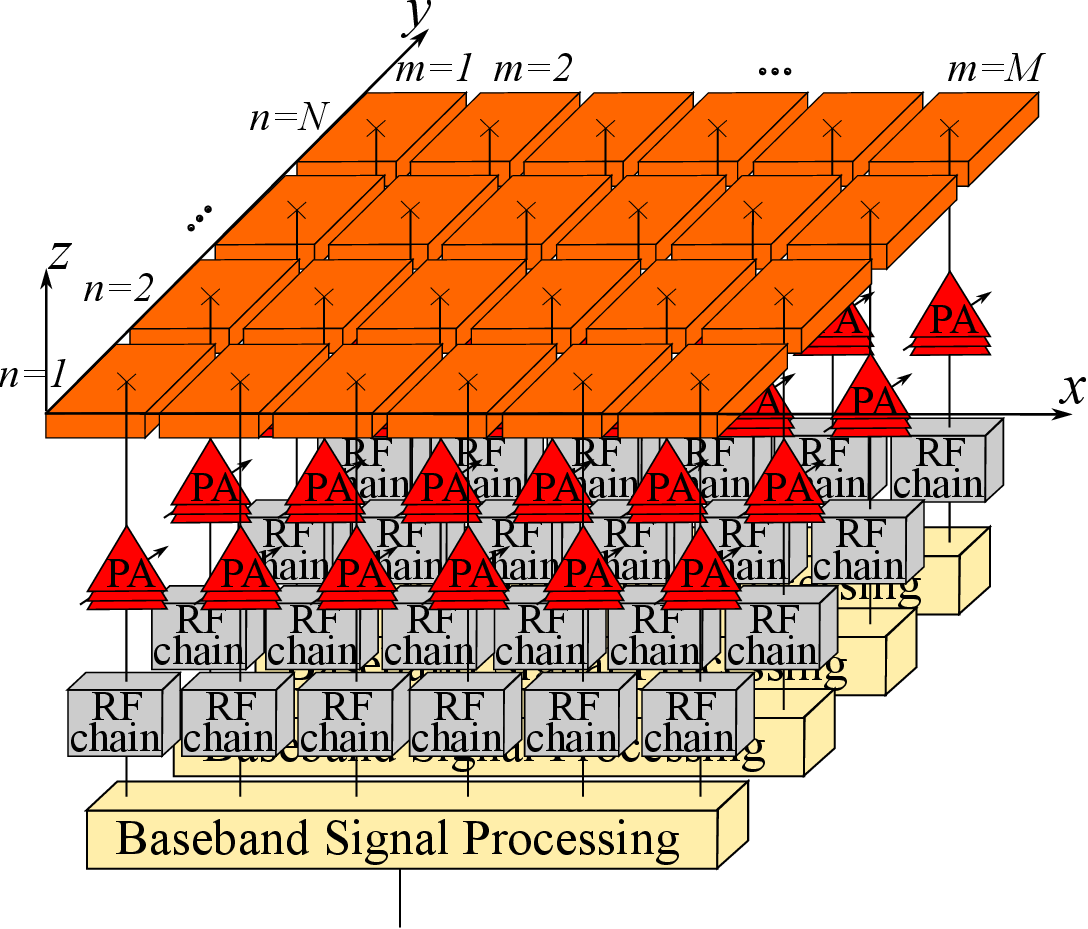}
    \includegraphics[width=0.245\linewidth]{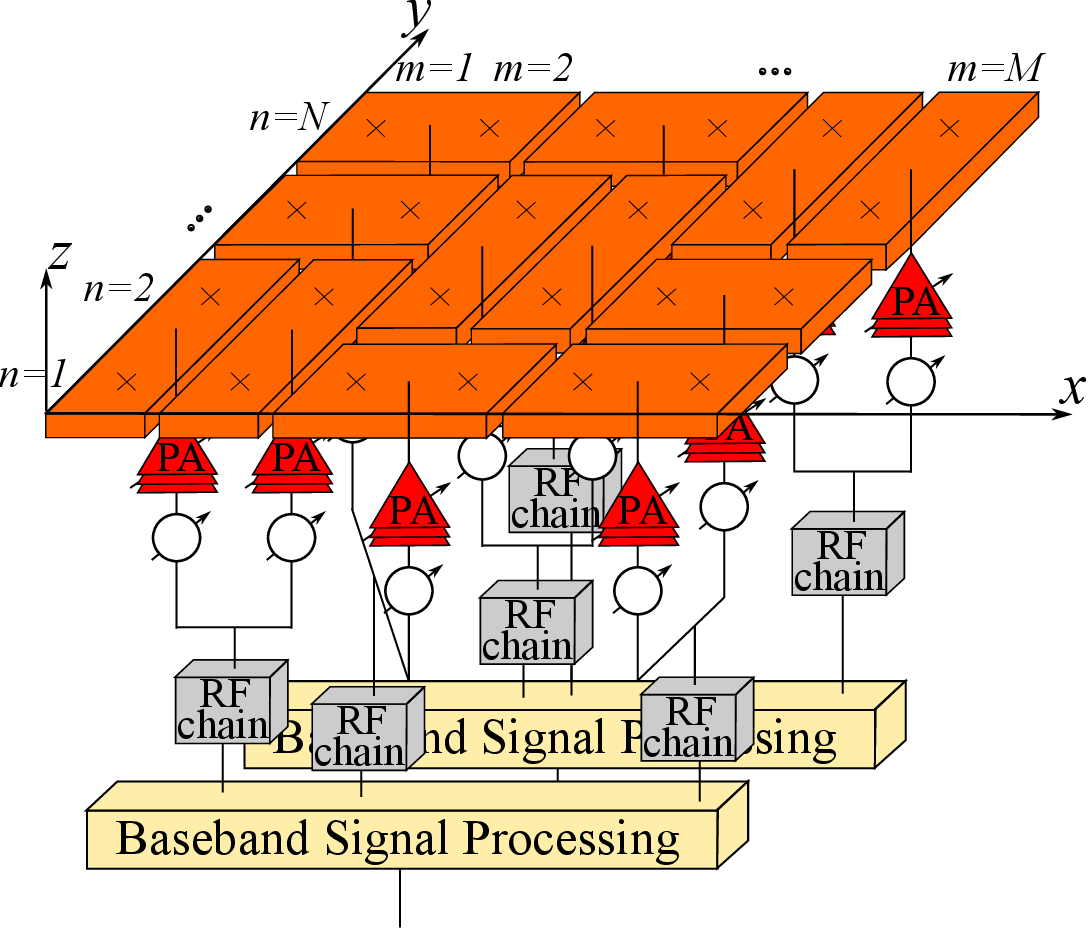}
    \includegraphics[width=0.245\linewidth]{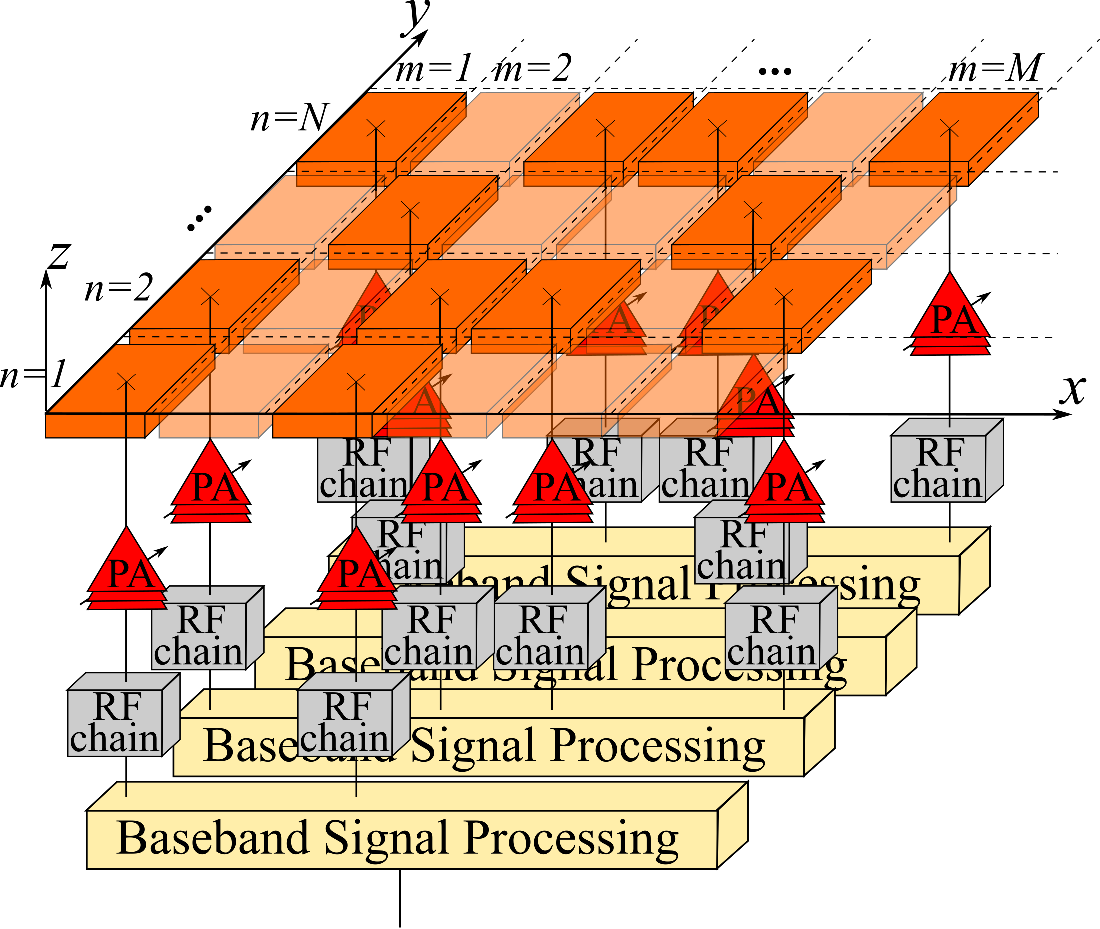}
    \includegraphics[width=0.245\linewidth]{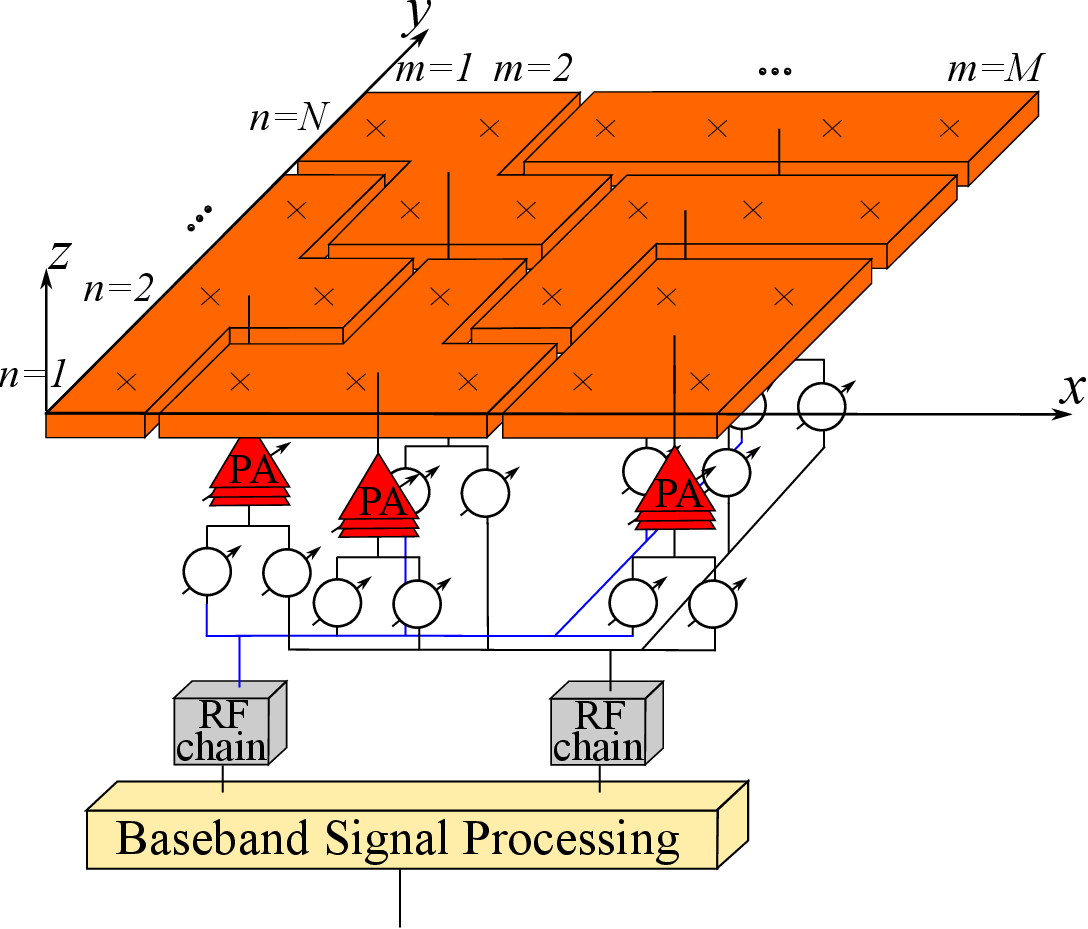}\\
    (a)\hspace{120pt}(b)\hspace{120pt}(c)\hspace{120pt}(d)\\
    \caption{Sketch of (a)~fully populated $N~\times~M~=~4~\times~6$\ reference array architecture~$(S~=~24)$, (b)~domino\--tiled phased array hybrid partially connected architecture~$(S~=~12)$, (c)~thinned phased array digital architecture~$(S~=~12)$, (d)~tetromino\--tiled phased array hybrid fully connected architecture~$(S~=~6)$.}    
    \label{fig:FPRA}
\end{figure*}

Consider a transmitter serving $K$ receivers, each equipped with $N_{RX}$ antennas oriented as $N'~\times~M'$, located at the far\--field of the transmitter. The transmit module comprises $N~\times~M~=~N_{TX}$ identical radiating elements, each excited by an independent feeding point where amplitude and/or phase control per antenna is possible [see Fig.~\ref{fig:FPRA}(a)]. In addition, each antenna has a dedicated RF chain, which allows full digital (FD) processing per antenna. Moreover, it is designed such that all the elements exhibit the same polarization. The symbols used in this work are listed in Table~\ref{ParamList}.
\begin{table}[h]
\caption{List of notations and symbols}\label{ParamList}%
\begin{tabular}{|cl|}
\hline
\textbf{Symbol} & \textbf{Definition} \\
\hline
$N$    & number of vertical transmit antenna elements   \\
$M$    & number of horizontal transmit antenna elements   \\
$N_{TX}$   & number of transmit antennas    \\
$N'$ & number of vertical receive antenna elements \\
$M'$ & number of horizontal receive antenna elements \\
$N_{RX}$ & number of receive antennas \\
$K$ & number of receivers \\
$B(u,v)$ & far\--field beam pattern radiated by the planar array\\
$\mathcal{D}$ & set of angles for SLL calculation \\
$g_{m,n}(u,v)$ & embedded/active element pattern of antenna $(m,n)$ \\
$x_{m,n}$ & element location in $x\--$axis \\
$y_{m,n}$ & element location in $y\--$axis \\
$a_{m,n}$   & complex array excitation coefficient  \\
$S$ & number of feeding points \\
$\mathbf{F}$ & precoding matrix \\
$\mathbf{P}$ & connection matrix between antennas and power amplifiers \\
$\mathbf{H}_k$ & channel between transmitter and receiver $k$ \\
$P_{PA}^{(out)}$ & power amplifier output power \\
$P_L$ & subarray feeding loss \\
$\mathbf{n}_o$ & additive noise \\
$\mathbf{r}_k$ & received signal at the antenna array of receiver $k$ \\
$\mathbf{w}$ & combining vector \\
$\eta$ & signal\--to\--noise ratio \\
$\gamma_k$ & signal\--to\--interference\--noise ratio at receiver $k$ \\
$R_k$ & SE of receiver $k$ \\
$\bar{R}$ & average sum spectral efficiency \\
$\Phi$ & SLL function \\
$\bar{R}_{ref}$ & sum SE of FPRA \\
$\Phi_{ref}$ & SLL of FPRA \\
$\beta$ & tuning parameter \\
$\mathcal{P}$ & set of thinned or clustered arrays \\
$\rho$ & fill factor of thinned array \\
$\mathcal{P}_{S}$ & set of $N~\times~M$ thinned array configurations, with $S$ feeds \\
$\mathcal{A}$ & set of TTPA subarrays \\
$g_{m,n}^{t}(u,v)$ & embedded/active element pattern of thinned array \\
$g_{m,n}^{\text{h,v}}(u,v)$ & embedded/active element pattern of DTPA \\
$g_{m,n}^{\mathcal{A}}(u,v)$ & embedded/active element pattern of TTPA \\
$\mathbf{Q}$ & dictionary matrix used in Algorithm\--X \\
$p_c$ & \textit{single-point} crossover probability \\
$p_m$ & \textit{single-point} mutation probability \\
$A$ & population size for GA search \\
$I_{max}$ & maximum number of iteration for GA search\\
$\chi$ & predefined performance for early termination \\
\hline
\end{tabular}
\end{table}
The far\--field beam pattern radiated~\cite{Balanis} by the planar array is given by
\begin{equation}
\label{eq:farfield}
    B(u,v) = \sum_{m=1}^M\sum_{n=1}^Ng_{m,n}(u,v)a_{m,n}e^{j2\pi\left(ux_{m,n}+vy_{m,n}\right)},
\end{equation}
where $g_{m,n}(u,v)$ is the embedded/active element pattern~\cite{Balanis}, $a_{m,n}~\in~\mathbb{C}$ is the complex excitation, $u~=~\sin\theta\cos\phi$, $v~=~\sin\theta\sin\phi$ and $x_{m,n}$ and $y_{m,n}$ are the element locations in wavelengths, $\theta~\in~[0,\frac{\pi}{2}]$ and $\phi~\in~[0,2\pi]$ are the spherical/angular coordinates, respectively.

Let $\mathbf{b}~=~\left[b_1,b_2,\ldots,b_K \right]~\in~\mathbb{C}^{K\times1}$ be the data to be transmitted to $K$ users satisfying $\mathbb{E}[\mathbf{bb}^H]~=~\mathbf{I}_K$. Prior to transmission, the array excitation coefficients $a_{m,n}$ in \eqref{eq:farfield} are computed utilizing a precoding matrix $\mathbf{F}~=~\left[\mathbf{f}_1,\mathbf{f}_2,\ldots,\mathbf{f}_K\right]~\in~\mathbb{C}^{S\times K}$ where $S$ is the number of feeding points $(S\leq N_{TX})$, i.e. $\mathbf{x}~=~\mathbf{Fb}$. The fully populated reference array (FPRA) architecture has $S~=~N_{TX}$ where each radiating patch can be controlled digitally. The precoded data $\mathbf{x}$ is amplified and mapped to antennas by a matrix $\mathbf{P}~\in~\mathbb{C}^{N_{TX}\times S}$. The mapping matrix is equal to identity for FPRA architecture, i.e. $\mathbf{P}~=~\mathbf{I}_S$, while it will be useful to represent hybrid and irregular array architectures [see Section~\ref{sec:HybArch} and Section~\ref{sec:IrrArrays}]. The amplitude of each unit element can be further tapered to control the SLL. The equivalent isotropically radiated power (EIRP) for data stream $k$~\cite{Ertugrul2023} can be calculated as 
\begin{equation}
    \text{EIRP}\left(u,v\right)_k=\frac{SP_{PA}^{(out)}}{KP_L}\left|B\left(u,v\right)\right|^2,
\end{equation}
where $P_{PA}^{(out)}$ is the output power of a single PA, $P_L$ is the feeding loss, and the beam pattern $B(u,v)$ is excited by the $k^{th}$ column of the precoder matrix $\mathbf{F}$, $a~=~\mathbf{f}_k$. The received signal at the antenna array of receiver $k$ can be written as
\begin{equation}
    \mathbf{r}_k=\mathbf{H}_k\mathbf{Pf}_kb_k + \sum_{j=1|j\neq k}^K\mathbf{H}_k\mathbf{Pf}_jb_j + \mathbf{n}_o,
\end{equation}
where $\mathbf{n}_o~\sim~\mathcal{CN}\left(0,\sigma^2\mathbf{I}_{N_{RX}}\right)$ is the additive noise and $\mathbf{H}_k~\in~\mathbb{C}^{N_{RX}\times N_{TX}}$ is the channel between the TX and RX $k$. In particular sub-THz channel model~\cite{Ju2021Indoor} has been considered
\begin{equation}
    \mathbf{H}_k = \sum_{\ell=1}^{L_k}\sum_{i=1}^{\kappa_{\ell}} \alpha_{\ell,i}\mathbf{a}_{RX}\left(u_{\ell,i}^{RX},v_{\ell,i}^{RX}\right)\mathbf{a}_{TX}^{T}\left(u_{\ell,i}^{TX},v_{\ell,i}^{TX} \right),
\end{equation}
where $L_k$ is the number of channel paths between the transmitter and the receiver $k$, $\kappa_{\ell}$ is the number of subpaths associated to path $\ell$, $\alpha_{\ell,i}~\in~\mathbb{C}$ is the complex channel gain for subpath $(\ell,i)$, $\left(u_{\ell,i}^{RX},v_{\ell,i}^{RX}\right)$ is the angle of arrival associated to subpath $i$ in path $\ell$, $\left(u_{\ell,i}^{TX},v_{\ell,i}^{TX}\right)$ is the angle of departure associated to subpath $i$ in path $\ell$, respectively.

The transmitter and receiver steering vectors are defined as follows,
\begin{multline}
    \mathbf{a}_{TX}(u,v) = [g_{1,1}(u,v)e^{j2\pi(ux_{1,1}+vy_{1,1})},\\ g_{1,2}(u,v)e^{j2\pi(ux_{1,2}+vy_{1,2})},
    \ldots,\\ g_{M,N}(u,v)e^{j2\pi(ux_{M,N}+vy_{M,N})}],
\end{multline}
\begin{multline}
    \mathbf{a}_{RX}(u,v) = [g_{1,1}(u,v)e^{j2\pi(ux'_{1,1}+vy'_{1,1})},\\ g_{1,2}(u,v)e^{j2\pi(ux'_{1,2}+vy'_{1,2})},
    \ldots,\\ g_{M',N'}(u,v)e^{j2\pi(ux'_{M',N'}+vy'_{M',N'})} ],
\end{multline}
where $(x'_{m',n'},y'_{m',n'})$ is the antenna element location for the receiver. Finally, the channel is normalized to have $\mathbb{E}[||\mathbf{H}_k||_F^2]~=~\bar{g}(u,v)$, where $\bar{g}(u,v)$ is the average embedded/active element pattern for transmitter and receiver antennas. 

Each receiver further processes the received signal by passing through the low\--noise amplifier and combining with vector $\mathbf{w}~\in~\mathbb{C}^{N_{RX}\times 1}$, i.e. $y_k~=~\mathbf{w}^H\mathbf{r}_k$. Note that combining vectors obey the unit norm, i.e. $\left|\left|\mathbf{w}^H\mathbf{w}\right|\right|^2~=~1$. The instantaneous signal\--to\--interference\--noise ratio (SINR) at receiver $k$ is denoted by $\gamma_k$, can be calculated as
\begin{equation}
    \gamma_k=\frac{\eta\left|\mathbf{w}^H\mathbf{H}_k\mathbf{P}\mathbf{f}_k\right|^2}{\eta\sum_{j=1|j\neq k}^K\left|\mathbf{w}^H\mathbf{H}_k\mathbf{P}\mathbf{f}_k\right|^2 + 1},
\end{equation}
where $\eta$ is the signal\--to\--noise ratio (SNR). The SE for the link between the transmitter and the receiver $k$ can be calculated by
\begin{equation}
    R_k=\log_2\left(1+\gamma_k\right)\quad \texttt{[b/s/Hz]}.
\end{equation}
The sum SE is $R~=~\sum_{k=1}^KR_k$ and the average SE can be calculated over many channel realizations, i.e., $\bar{R}~=~\mathbb{E}[R]$. The SLL of a beam pattern is denoted by $\Phi~=~\text{SLL}\left[|B(u,v)|^2\right]$, which is a function of the amplitude excitation $a$ that is computed by a collection of channel realizations for all receivers, i.e., $\mathbf{H}_1,\mathbf{H}_2,\ldots,\mathbf{H}_K$. In particular, the SLL can be expressed as the maximum radiation level that is outside the main beam(s)
\begin{equation}
    \Phi = \max_{(u,v)\in \mathcal{D}} |B(u,v)|^2,
\end{equation}
where the set $\mathcal{D}$ contains all $(u,v)$ directions that is outside the main beam(s), i.e. $\mathcal{D}~=~\bigcap_{k=1}^K\left\{|u-u_k|\geq \bar{u} \right\}~\cup~\left\{|v-v_k|\geq\bar{v} \right\}$, $(u_k,v_k)$ stands for the direction of the main beam towards receiver $k$, $\bar{u}$ and $\bar{v}$ stand for the angular spread of the main beam in $u-$ and $v-$ planes, respectively.
\subsection{Hybrid Architectures}\label{sec:HybArch}
Hybrid architectures are realized by factorizing the precoder matrix $\mathbf{F}$ into an analog part, $\mathbf{F}_{RF}~=~\left[\mathbf{f}_{RF,1},\mathbf{f}_{RF,2},\ldots,\mathbf{f}_{RF,K}\right]\in\mathbb{C}^{S\times K}$ and a baseband part, $\mathbf{F}_{BB}\in\mathbb{C}^{K\times K}$, resulting in 
$\mathbf{F}=\mathbf{F}_{RF}\mathbf{F}_{BB}$. Specifically, the non-zero entries in the RF analog precoder, are represented as phase shifts, and a unit modulus constraint is imposed on them. 

Two types of hybrid architectures are investigated: hybrid fully connected (HFC) [see Fig.~\ref{fig:FPRA}(d)] and hybrid partially connected (HPC) [see Fig.~\ref{fig:FPRA}(b)]. In the HFC architecture, all RF chains are connected to each feeding point, while in the HPC architecture, each RF chain is connected to a disjoint subset of feeding points. Therefore, the RF precoding matrix for the HFC architecture is fully populated, i.e., $f_{RF}(s,k)=\frac{1}{\sqrt{S}}e^{j\varphi_{s,k}}$, whereas a block-diagonal structure is realized for the HPC architecture, with $f_{RF}(s,k)=\sqrt{\frac{K}{S}}e^{j\varphi_{s,k}}$.

A two-step hybrid precoding approach~\cite{Alkhateeb2015} is adopted in this work. The RF precoding matrix $\mathbf{F}_{RF}$ is first determined, followed by the computation of the baseband precoder $\mathbf{F}_{BB}$  based on the effective channel seen by the RF chains~\cite{Alkhateeb2015}. The RF precoding matrix is computed as follows:
\begin{equation}
    \left\{\mathbf{w}_{RF,k}^{\star}, \mathbf{f}_{RF,k}^{\star}\right\} = \arg\max\left|\mathbf{w}_{RF,k}^H\mathbf{H}_k\mathbf{P}\mathbf{f}_{RF,k} \right|^2,
\end{equation}
where $\mathbf{w}_{RF,k}\in\mathcal{W}$ and $\mathbf{f}_{RF,k}\in\mathcal{F}$, $\mathcal{W}$ and $\mathcal{F}$ are the receiver and transmitter RF codebooks which search takes place, respectively. The effective channel seen by the digital baseband can be expressed as $\bar{\mathbf{H}}_k=\mathbf{w}_{RF,k}\mathbf{H}_k\mathbf{P}\mathbf{f}_{RF,k}$. In the second step, a zero forcing (ZF) precoder~\cite{Spencer2004} is employed on the aggregated effective channel matrix $\bar{\mathbf{H}}=\left[\bar{\mathbf{H}}_1^T,\bar{\mathbf{H}}_2^T,\ldots,\bar{\mathbf{H}}_K^T \right]^T$. Hence, the baseband precoder can be computed as
\begin{equation}
    \mathbf{F}_{BB} = \bar{\mathbf{H}}^H\left(\bar{\mathbf{H}}\bar{\mathbf{H}}^H \right)^{-1}.
\end{equation}

\subsection{Irregular Arrays}\label{sec:IrrArrays}
To assess the performance of the thinned and clustered arrays, the signal model presented in the previous section is modified. The connection matrix $\mathbf{P}$ indicates the irregular array configuration. Thus, we are interested in the following objective function,
\begin{equation}
\label{eq:OptProb}
    \mathbf{P}^{\star} = \arg\max_{\mathbf{P}\in\mathcal{P}}\beta\frac{\bar{R}\left(\mathbf{P}\right)}{\bar{R}_{ref}} + \left(1-\beta\right)\frac{\Phi\left(\mathbf{P}\right)}{\Phi_{ref}},
\end{equation}
where $\bar{R}_{ref}$ and $\Phi_{ref}$ stand for the sum SE and SLL corresponding to an FPRA. The optimization takes place among all possible irregular array configurations $\mathcal{P}$ with the desired property. In addition, $\beta\in[0,1]$ is a parameter that gives priority to communications performance or SLLs. In particular, $\beta=1$ corresponds to pure communications performance while $\beta=0$ corresponds to pure SLL optimization, respectively.
\subsubsection{Thinned Array}\label{sec:ThinArray}
Removal of the radiating elements from a FPRA architecture results in a thinned array [see Fig.~\ref{fig:FPRA}(c)] which can be modeled by modifying the connection matrix $\mathbf{P}$. Zero rows are inserted in the locations where there is no connection, The remaining rows contain ones only in one location each, where the antenna $j$ is connected to PA $i$ indicated as $\mathbf{P}(i,j)=1$. The embedded/active element pattern $g_n(u,v)$ is modified to compute the far\--field response of the thinned array as
\begin{equation}
    g_n(u,v)=g_n^{\text{t}}(u,v),
\end{equation}
where $g_n^{\text{t}}(u,v)$ is the embedded/active element pattern for the thinned array. The ratio of the non\--zero rows to the total number of antennas is indicated by the fill factor
\begin{equation}
    \rho=\frac{S}{N_{TX}}.
\end{equation}
We denote the solution space for an $N~\times~M$ thinned array with $S$ elements by $\mathcal{P}_S$ and counted as follows,
\begin{multline}
\label{eq:ThinSoln}
    |\mathcal{P}_S|= \begin{pmatrix}
        N_{TX}\\ S
    \end{pmatrix} - 2\begin{pmatrix}
        N_{TX}-M \\ S
    \end{pmatrix}
    -2\begin{pmatrix}
        N_{TX} - N \\ S
    \end{pmatrix}\\
    + 4\begin{pmatrix}
       N_{TX}-N-M+1 \\ S 
    \end{pmatrix},
\end{multline}
where the last three terms exclude the thinned arrays with smaller apertures than $N~\times~M$.
We generate random samples based on the method in~\cite{Zhang2023} to search the solution space $\mathcal{P}_S$. Later, samples that maximize the objective~\eqref{eq:OptProb} are selected with respect to $\beta$.
\subsubsection{Clustered Array}\label{sec:ClusArray}
Clustered arrays are constructed by grouping the neighboring radiating elements, which are denoted as tiles, having domino [see Fig.~\ref{fig:FPRA}(b)] and tetromino [see Fig.~\ref{fig:FPRA}(d)] shapes and grouping of two and four neighboring antenna elements, respectively. Domino tile shapes can be realized in vertical and horizontal orientations that combine two adjacent unit elements sharing a horizontal or a vertical edge, respectively. In the case of tetromino, there are 19 possible tile shapes, including rotations and reflections. The element pattern $g_{m,n}(u,v)$ and element feed locations $x_{m,n}$ and $y_{m,n}$ are modified to compute the far\--field radiation pattern of the domino\--tiled phased arrays (DTPAs) and tetromino\--tiled phased arrays (TTPAs), respectively. 
\begin{equation}
    g_{m,n}(u,v)=
    \begin{cases}
        g_{m,n}^{\text{v}}(u,v)\quad \text{vertical domino}\\
        g_{m,n}^{\text{h}}(u,v)\quad \text{horizontal domino},
    \end{cases}
\end{equation}
for the DTPAs and 
\begin{equation}
    g_{m,n}(u,v)=g_{m,n}^{\mathcal{A}}(u,v),
\end{equation}
for the TTPAs where $\mathcal{A}$ stands for the set of 19 possible tetromino unit elements, including rotations and reflections. The antenna locations are updated by replacing the coordinates of the antenna feeding points $\left(x_{m,n},y_{m,n} \right)$ by $\left(\bar{x}_{m,n},\bar{y}_{m,n} \right)$ which can be computed as 
\begin{equation}
    \bar{x}_{m,n}=\frac{x_{m,n}+x_{m,n}'}{2}
\end{equation}
and
\begin{equation}
    \bar{y}_{m,n}=\frac{y_{m,n}+y_{m,n}'}{2}
\end{equation}
for DTPAs where  $\left(x_{m,n},y_{m,n} \right)$ and  $\left(x_{m,n}',y_{m,n}' \right)$ are the coordinates of the neighboring antenna elements sharing the same tile. Similarly, the antenna locations are updated by
\begin{equation}
    \bar{x}_{m,n}=\frac{x_{m,n}+x_{m,n}'+x_{m,n}''+x_{m,n}'''}{4}
\end{equation}
and
\begin{equation}
    \bar{y}_{m,n}=\frac{y_{m,n}+y_{m,n}'+y_{m,n}''+y_{m,n}'''}{4}
\end{equation}
for TTPAs where $\left(x_{m,n},y_{m,n}\right)$, $\left(x_{m,n}',y_{m,n}'\right)$, $\left(x_{m,n}'',y_{m,n}''\right)$, and $\left(x_{m,n}''',y_{m,n}'''\right)$ are neighboring antenna element coordinates sharing the same tile. Furthermore, the connection matrix $\mathbf{P}$ can be modified to reflect the behavior of the clustered arrays. In particular, for DTPAs the connection matrix $\mathbf{P}$ has a column sum of 2, indicating a PA connecting to 2 radiating elements [see Fig.~\ref{fig:FPRA}(b)]. Similarly, for TTPAs, the connection matrix $\mathbf{P}$ has a column sum of 4, indicating a PA connecting to 4 radiating elements [see Fig.~\ref{fig:FPRA}(d)]. The size of the solution space $|\mathcal{P}|$ of DTPA for an $N~\times~M$ aperture can be calculated as~\cite{KASTELEYN1961}
\begin{equation}
\label{eq:DominoSoln}
    |\mathcal{P}|=2^{\frac{NM}{2}}\prod_{m=1}^M\prod_{n=1}^{N}\left[\cos^2\left(\frac{\pi m}{M+1}\right) + \cos^2\left(\frac{\pi n}{N+1}\right)\right]^{1/4}.
\end{equation}
The enumerative tiling method (ETM) described in~\cite{Anselmi2017} is used to generate many DTPA samples and a GA-based search tailored to find optimal DTPAs~\cite{Anselmi2017} considering the objective~\eqref{eq:OptProb}.
We use algorithm\--X~\cite{knuth2000} to generate TTPA configurations that cover the $N~\times~M$ planar array. Among many generated samples, we utilize GA search~\cite{Johnson1997} to find the optimal array configuration. The binary representation of the sample index is used as a chromosome in GA search, where $\lceil\log_2\left(|\mathcal{P}|\right)\rceil$ is the length of the chromosomes.

\section{Irregular Array Configuration Generation Methods}
\label{sec:ConfGen}
Configuration generation methods mentioned in Section~\ref{sec:IrrArrays} are explained in this part. GA\--like random search methods are flexible in terms of objective functions, i.e., they can work with non\--convex objectives and constraints but do not guarantee convergence to the global optimum. Hence, brute force search methodologies such as Algorithm\--X are explored to assess the performance of GA based search methods and to identify further statistics about the solution space.
\subsection {Algorithm-X}
\label{sec:algX}
\begin{figure}
    \centering
    \includegraphics[width=0.6\linewidth]{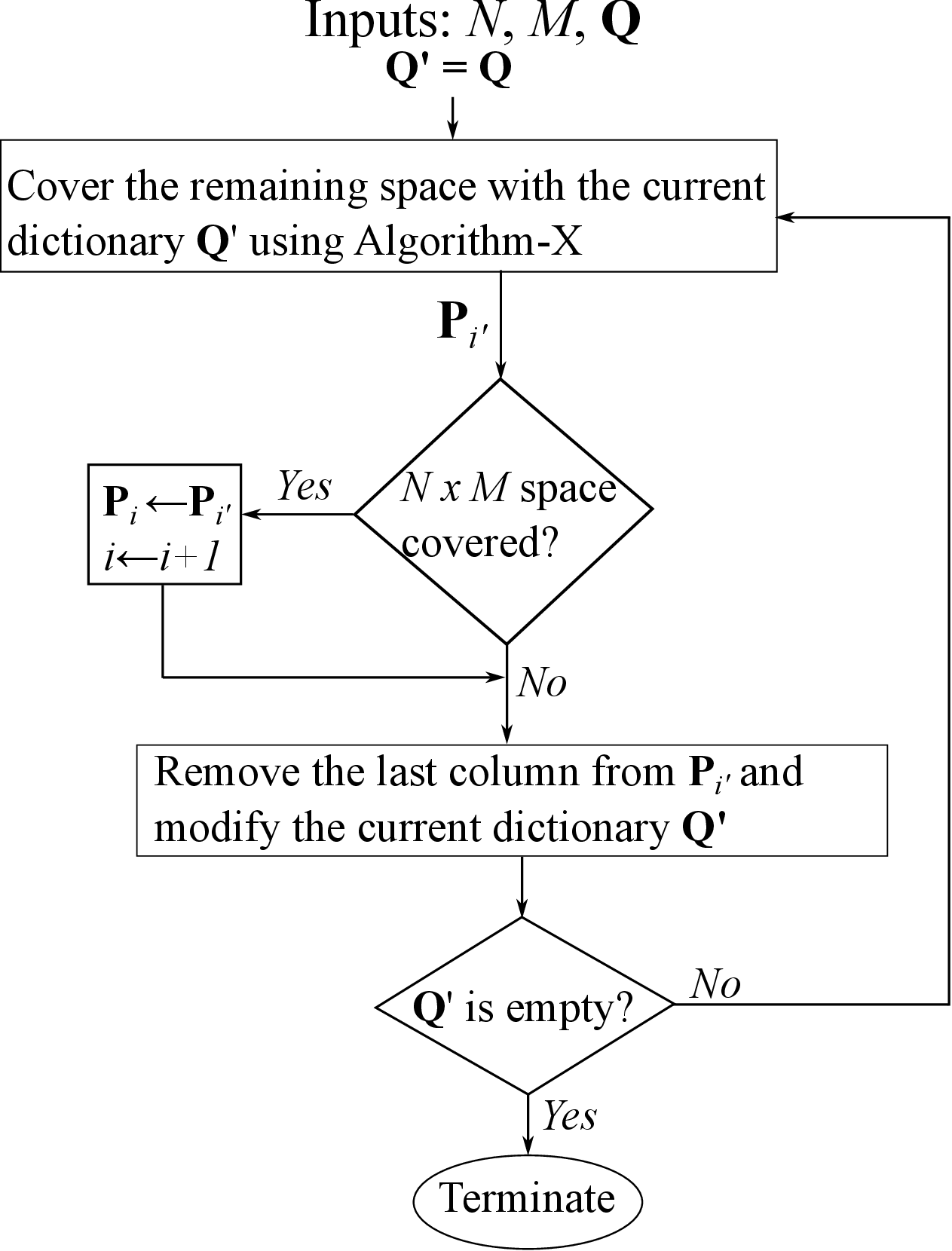}
    \caption{Algorithm\--X based irregular array configuration generation scheme.}
    \label{fig:AlgX}
\end{figure}

The initial version of the Algorithm\--X~\cite{knuth2000} is designed to find a configuration that covers the area of interest with the given unit pieces, also known as the exact cover problem. This algorithm can be modified to generate all possible irregular array configurations either in a brute\--force or a randomized fashion~\cite{Miao2022,Xiong2013}. In this work, we adopt the brute\--force sample generation in order to search the solution space for clustered arrays [see Section~\ref{sec:ClusArray}]. Hence, one needs to create a dictionary matrix $\mathbf{Q}$ that contains all possible elements and their configurations in $N~\times~M$ aperture. For example, a dictionary matrix for DTPA contains all possible arrangements of horizontal and vertical elements in $N~\times~M$ aperture. Fig.~\ref{fig:AlgX} depicts the flowchart for the brute force irregular array configuration generation scheme, which uses Algorithm\--X and manipulates the dictionary matrix $\mathbf{Q}$. The algorithm updates the current dictionary $\mathbf{Q}'$ which is nothing but a subset of the dictionary matrix $\mathbf{Q}$ and updates the last configuration $\mathbf{P}_i$ that covers $N~\times~M$ aperture by by removing the recent elements that has been placed in $\mathbf{P}_{i'}$. The subscript $i$ in $\mathbf{P}_i$ stands for the configuration index used for enumerating the generated configurations.
\subsection {Genetic Algorithm}
\label{sec:GAalg}
GA has been used to search the solution space $\mathcal{P}$ effectively. First, the plain version of the GA is described, and necessary modifications are discussed for each irregular array architecture. One needs to come up with a unique representation of each sample that serves as a chromosome description. The algorithm runs for a maximum of $I_{max}$ iterations or stops when a predefined performance metric $\chi$ is satisfied.

\textit{1) Step 1 (Initialization):} An initial population of size $A$ is generated where each member of the population is a distinct configuration in the solution space $\mathcal{P}$. The fitness value of each sample is evaluated and enumerated according in the ascending order.

\textit{2) Step 2 (Reproduction Cycle):} Two new samples are generated, first by selecting parent samples with \textit{roulette\--wheel}~\cite{Johnson1997} selection, and combining them with the \textit{single\--point} crossover~\cite{Johnson1997} which can occur with probability $p_c$. Furthermore, \textit{single\--point} mutation~\cite{Johnson1997} modifies the newly generated samples individually, which can occur with probability $p_m$.

\textit{3) Step 3 (Fitness Evaluation):} Fitness values for the newly generated samples are computed. Finally, the \textit{elitism} operator~\cite{Johnson1997} is applied to assure that the best solution found so far stays in the population.

\textit{4) Step 4 (Convergence Check):} The algorithm is terminated if $I_{max}$ is reached or a solution is found that satisfies a user\--defined threshold $\chi$. Otherwise, the iteration number is incremented and \textit{Step 2} is executed.

\begin{figure}
    \centering
    \includegraphics[width=0.8\linewidth]{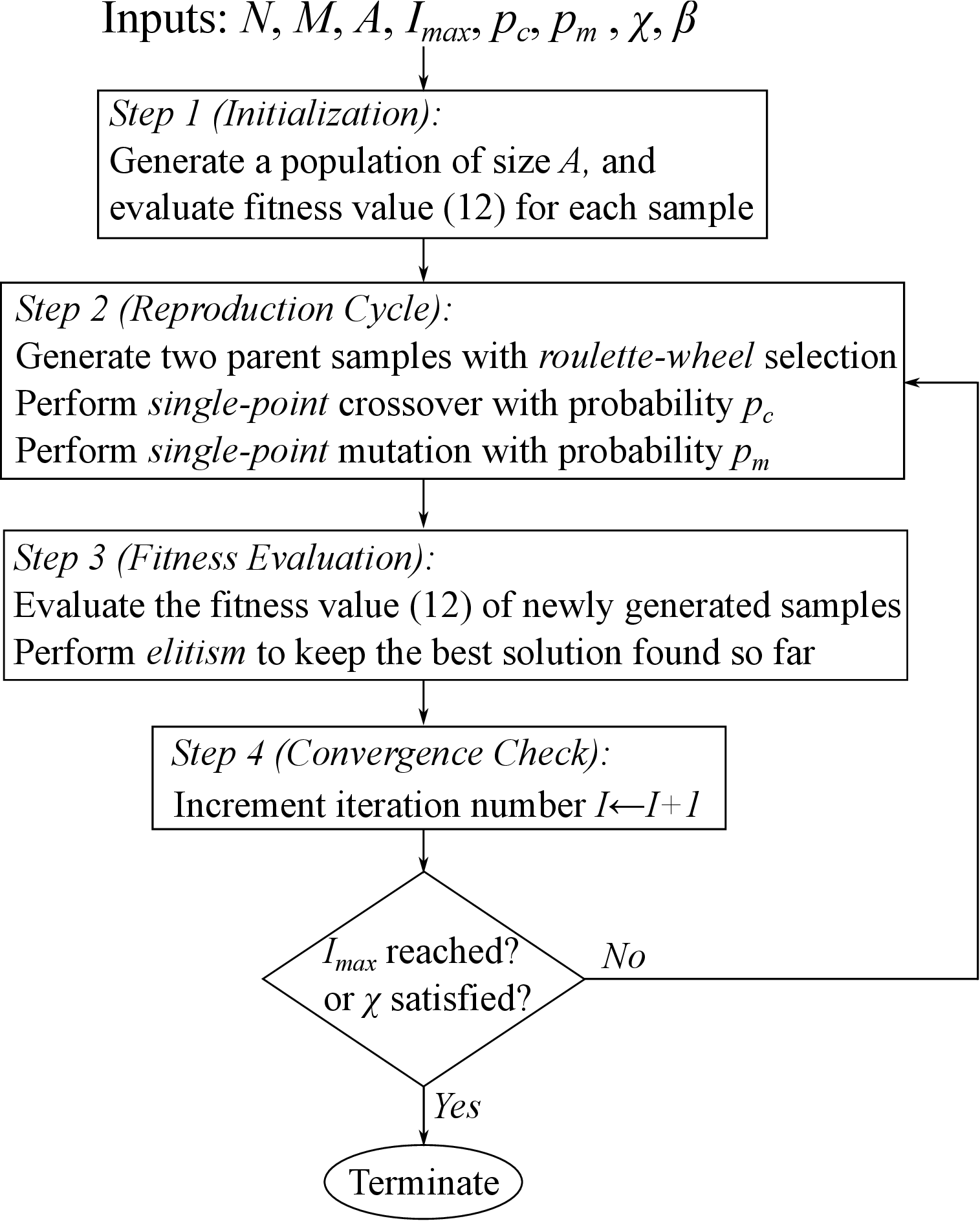}
    \caption{GA based irregular array optimization scheme.}
    \label{fig:GAflow}
\end{figure}

A flowchart explaining the GA based irregular array performance optimization is depicted in Fig.~\ref{fig:GAflow}.

The chromosome definition for DTPA follows the description in~\cite{Anselmi2017} that is inspired by the \textit{height function} in~\cite{Desreux2006}. In the case of TTPA configurations, we use the sample index that is determined during the configuration generation scheme described in Section~\ref{sec:algX} in binary format as a chromosome description similar to~\cite{Ma2024}.

\section{D-Band Antenna Array Design}\label{sec:CST}

\begin{figure}[]
    \centering
    \includegraphics[width=0.65\linewidth]{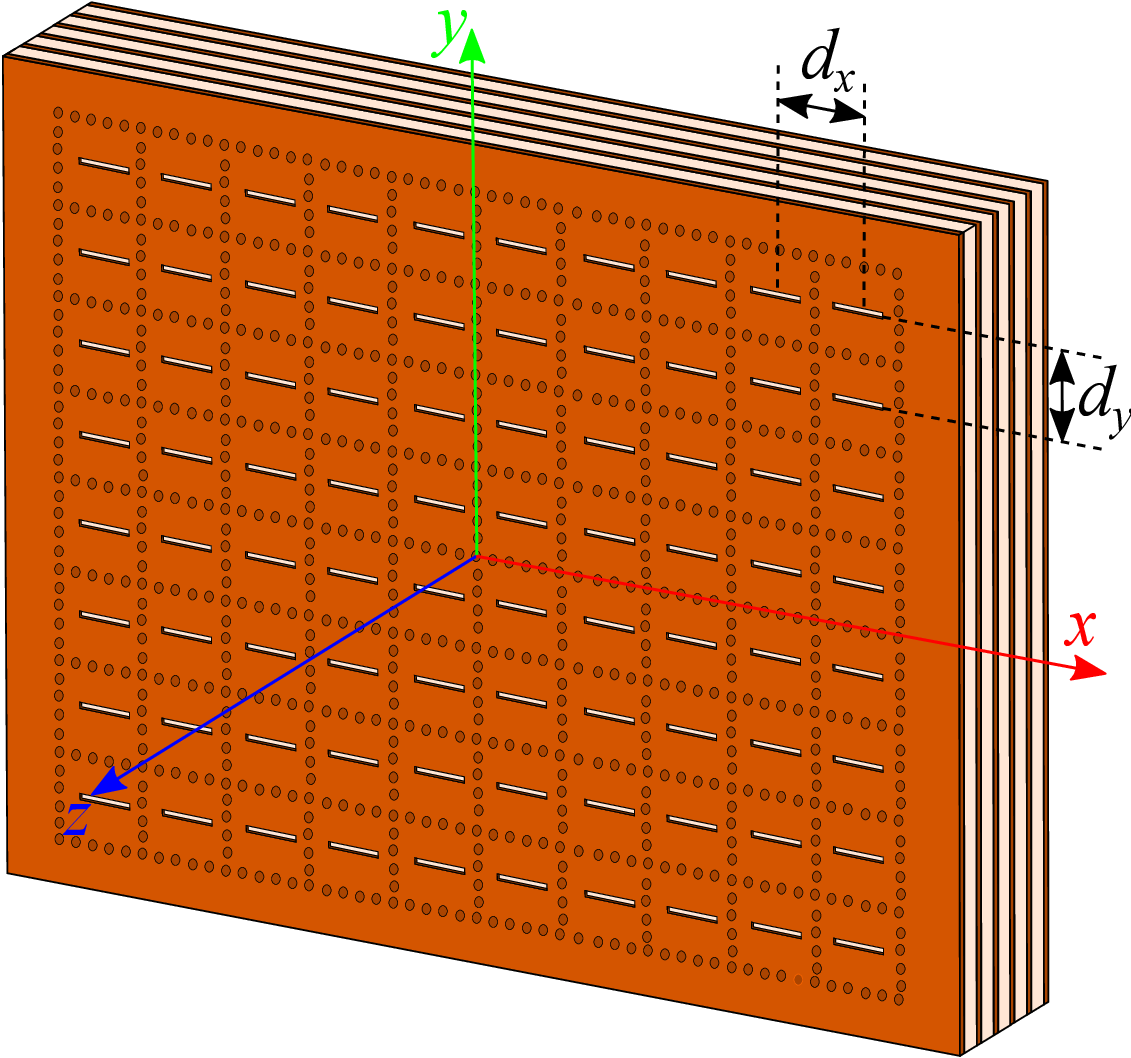}\\
    \small(a)\\
    \includegraphics[width=0.85\linewidth]{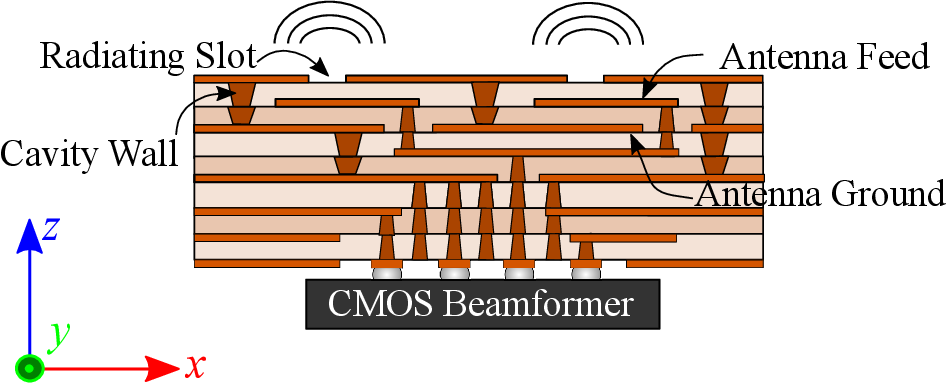}\\   
    \small(b)\\
    \caption{(a)~Layout diagram of the proposed 8$\times$10 array ($d_x$~=~$d_y$~=~0.5~$\lambda_0$). (b)~8-layer PCB stackup.}
    \label{fig:Array}
\end{figure}

A highly integrated, linearly polarized 8$\times$10 phased arrays with interelement spacings of $d_x$ and $d_y$ along the $x$- and $y$-axes, respectively, are designed on a printed circuit board (PCB) using the frequency-domain solver of CST Microwave Studio~[Fig.~\ref{fig:Array}(a)]. These arrays are developed to validate the performance of the synthesized arrays. Operating at 140~GHz, each antenna element is designed to fit within a compact footprint of less than 1.1mm$\times$1.1mm (0.5~$\lambda_0$$\times$0.5~$\lambda_0$).

\begin{figure}[]
    \centering
    \includegraphics[width=0.35\linewidth]{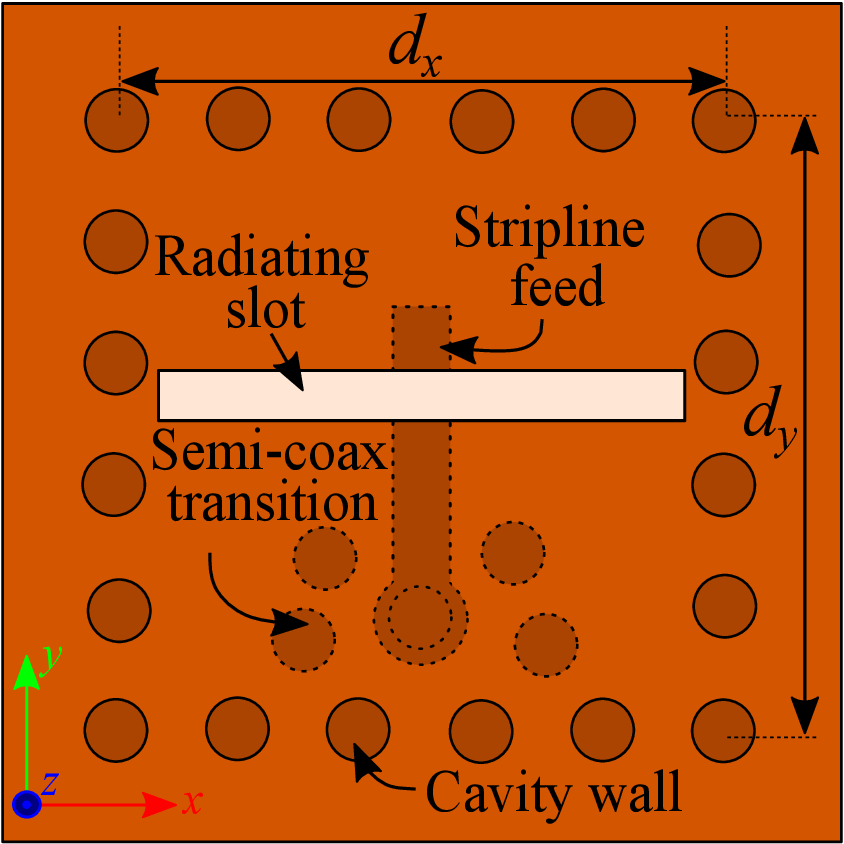}
     \includegraphics[width=0.45\linewidth]{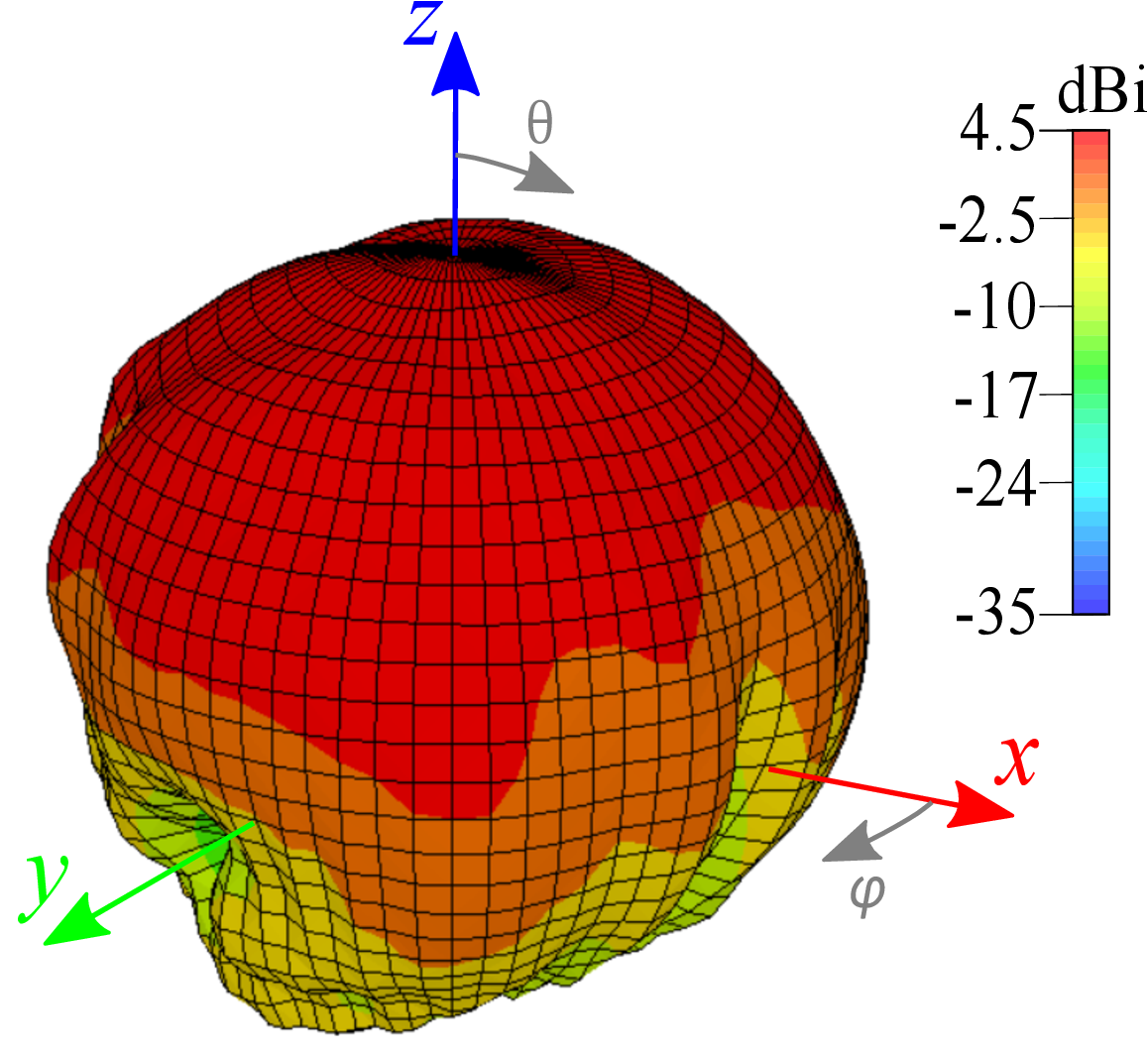}\\
    \small(a) \hspace{120pt}\small(b)\\
    \caption{(a)~Strip line feed cavity-backed slot antenna element. (b)~3D plot of the embedded gain pattern of the strip line feed cavity-backed slot arrays resonating at 140~GHz.}
    \label{fig:Pattern}
\end{figure}

The design process incorporates standard any-layer high-density interconnect (HDI) PCB constraints, ensuring seamless CMOS beamformer-to-antenna integration and facilitating mass production. To meet stringent performance requirements, a stripline-fed, cavity-backed slot antenna topology is adopted, as shown in Fig.~\ref{fig:Pattern}(a). The backing cavity ensures robust isolation between the antenna and the integration platform, while also minimizing mutual coupling in the phased array~\cite{Kapusuz2017}. Additionally, the rectangular slot configuration provides high co-to-cross polarization discrimination.

As illustrated in Fig.~\ref{fig:Array}(b), the CMOS beamformer~\cite{Zhang2024_2} is directly integrated into the tiled and tinned array configurations. The antenna is initially realized on a high-frequency PCB laminate, further optimizing its performance. To ensure high system efficiency, a dedicated stripline-to-coax transition is utilized to feed the antenna arrays. In order to control impedance during the transition, a semi-coaxial type structure is employed. In this configuration, one side of the vias functions as the outer conductor of the coaxial probe, while the other side serves as the inner conductor, connecting the antenna feed points and the stripline transmission line to the beamformer.

The simulated embedded radiation pattern of the antenna at 140~GHz is depicted in Fig.~\ref{fig:Pattern}(b). A simulated embedded boresight gain of 4.5~dBi is achieved at 140~GHz.

\section{Results \& Discussion}\label{sec:Results}

The system model developed in Section~\ref{sec:Sysmodel} has been validated through numerical simulations in this section. The performance of irregular arrays is evaluated in an indoor environment, where two receivers $(K~=~2)$, each equipped with a 
$4~\times~4$ square array $(N_{RX}~=~16)$, are positioned in the far-field of an 80-element rectangular array $N~\times~M~=~8~\times~10$ (see Section~\ref{sec:CST}), as described in~\cite{Ertugrul2024,Ertugrul2024_2}. For comparison, thinned and clustered array designs have been optimized for both the SLL and SE at 140~GHz, based on the objective function in~\eqref{eq:OptProb}.

The angular spread of the main beam is set at $\bar{u}~=~0.21$ and $\bar{v}~=~0.28$ in $\mathcal{D}$ for SLL evaluation. In a MU-MIMO configuration under the 140~GHz propagation environment~\cite{Ju2021Indoor}, it is expected that $K$ beams will form, targeting the receivers due to the sparse nature of the channel. The SLL and SE values are averaged over 25~channel realizations, with receivers randomly distributed on an angular grid.

The solution space sizes $|\mathcal{P}|$ for DTPA, as well as 25\% and 50\% thinned arrays, are presented in Table~\ref{Solnspace}, calculated using~\eqref{eq:ThinSoln}, and \eqref{eq:DominoSoln}. For an $8~\times~10$ array the size of the solution space for DTPA is around $10^9$ samples, thinned array with $\rho~=~0.5$ is around $10^{23}$, and thinned array with $\rho~=~0.25$ is around $10^{18}$, respectively.

\begin{table}[h]
\caption{The size of the solution space $|\mathcal{P}|$ for different irregular array architectures}\label{Solnspace}
\centering
\begin{tabular}{|clll|}
\hline
$N~=~M$ & DTPA  & 25\% Thinned & 50\% Thinned\\
\hline
6    & 6728   & $4.5\times 10^7$   & $8.7\times 10^9$  \\
8    & $1.29\times 10^7$   &$3.3\times 10^{14}$  & $1.8\times 10^{18}$  \\
10   & $2.58\times 10^{11}$   & $1.9\times 10^{23}$  & $10^{29}$  \\
16   & $2.44\times 10^{30}$    & $1.8\times 10^{61}$   & $5.7\times 10^{75}$  \\
\hline
\end{tabular}
\end{table}

Consequently, for the $8~\times~10$ transmit array, the size of the solution space is too large to be explored through brute-force search, making it necessary to apply the search methodologies outlined in Sections~\ref{sec:IrrArrays} and \ref{sec:ConfGen}. Hence, GA based search methodology described in Section~\ref{sec:GAalg} is utilized to effectively find the array configurations that maximize the objective \eqref{eq:OptProb}. We consider maximum number of iterations $I_{max}~=~300$ for DTPA optimization due to fast convergence thanks to the chromosome description~\cite{Anselmi2017} and $I_{max}~=~1000$ is selected for TTPA optimization while no predefined performance $\chi$ is used for early termination. In all GA runs, a population size of $A~=~20$ along with cross-over probability $p_c=0.9$, mutation probability $p_m=0.1$, and different discrete tuning parameter $\beta$ values are considered, respectively. For TTPA optimization, $2.5\times10^7$ irregular array configurations were generated, encoded as 25-bit binary vectors, that is used as chromosomes in the GA search. Thinned array configurations are optimized considering the technique in~\cite{Zhang2023}.

The $8~\times~10$ antenna array designed in Section~\ref{sec:CST} is employed for performance comparison. Without loss of generality, the unit element responses for the reference and thinned arrays at the broadside angle are assumed to be $g_{m,n}(0,0)~\simeq~4.07$~dBi and $g_{m,n}^{\text{t}}(0,0)~\simeq~5.68$~dBi, respectively, with no feeding losses, i.e. $P_L~=~0$~dB. Additionally, the unit element responses for DTPAs and TTPAs at the broadside angle are assumed to be $g_{m,n}^{\text{h,v}}(0,0)~\simeq~6.5$~dBi and $g_{m,n}^{\mathcal{A}}(0,0)~\simeq~7.9$~dBi, respectively, with feeding losses of $P_L~=~0.3$ dB and $P_L~=~0.6$ dB, respectively.

\begin{figure}[]
    \centering
    \includegraphics[width=0.48\linewidth]{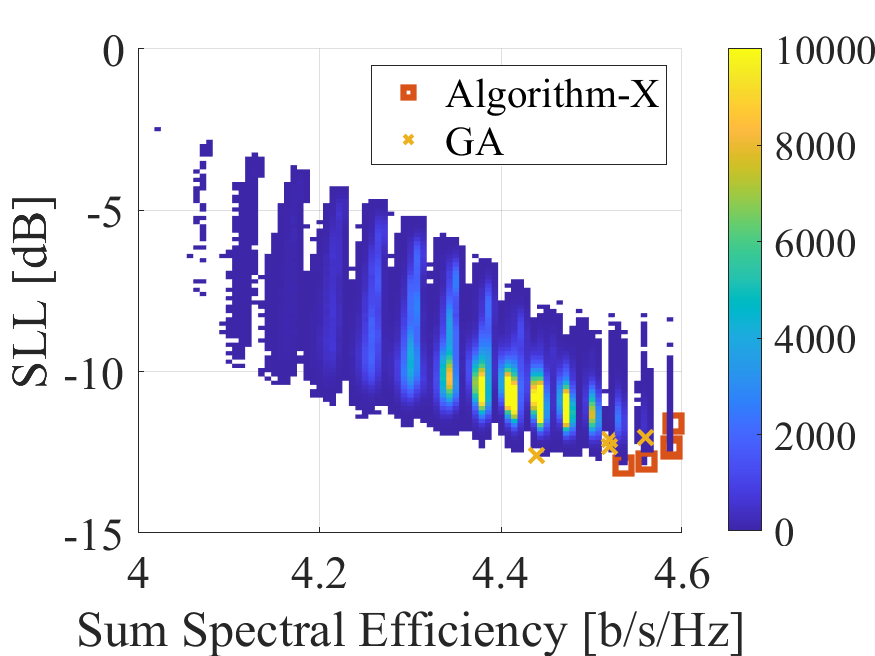}
    \includegraphics[width=0.48\linewidth]{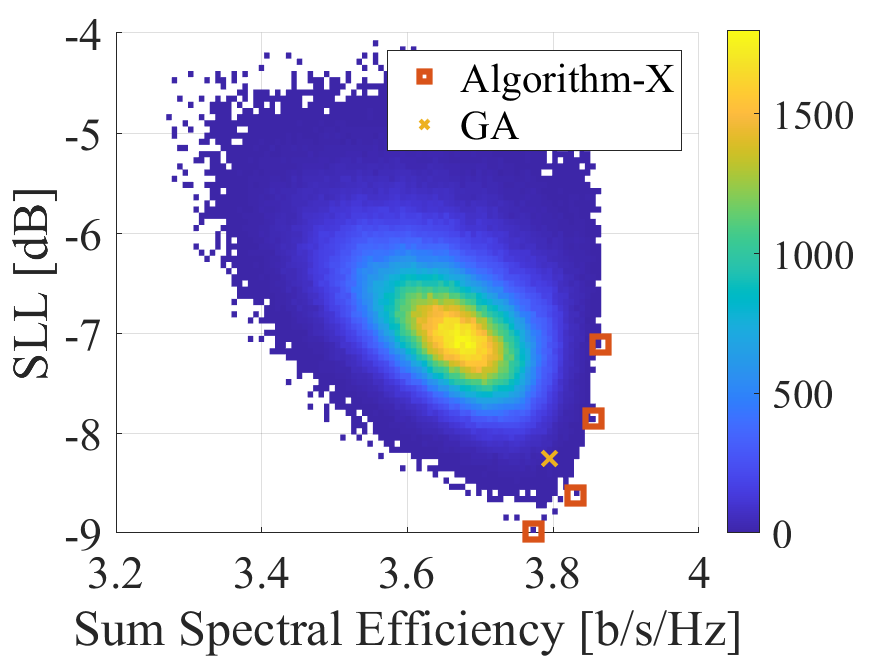}\\
    \small(a)\hspace{110pt}(b)\\
    \includegraphics[width=0.48\linewidth]{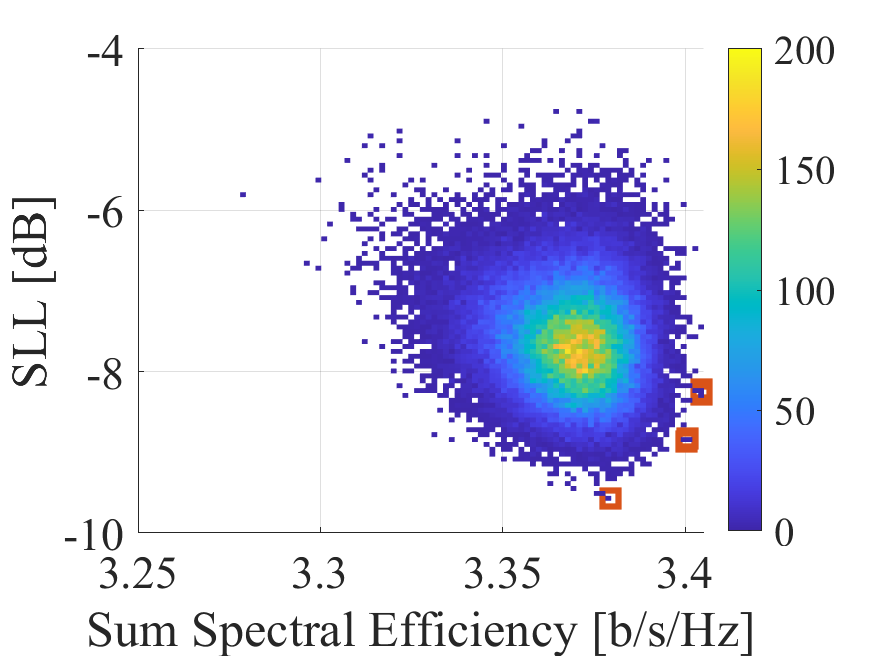}
    \includegraphics[width=0.48\linewidth]{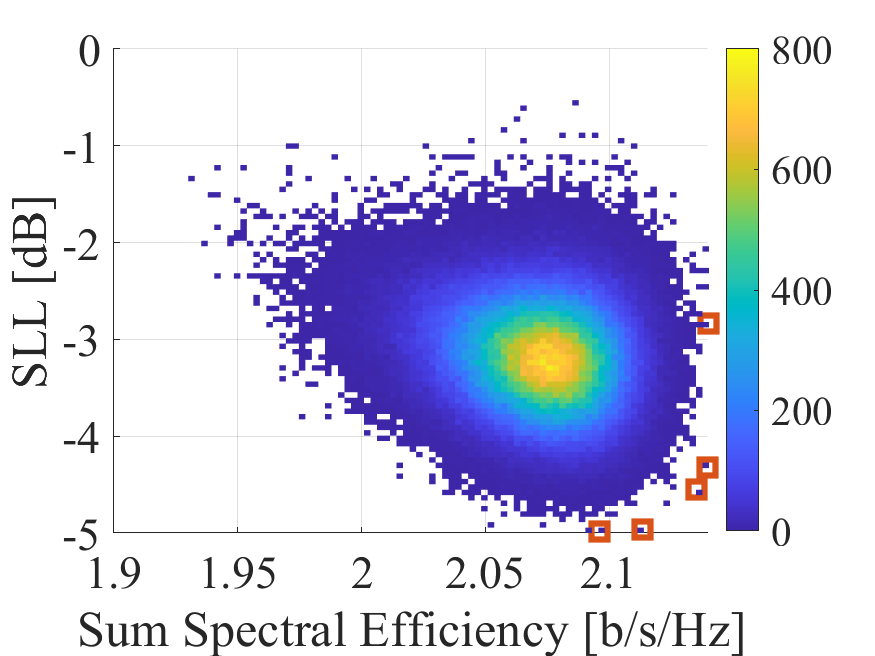}\\
    (c)\hspace{110pt}(d)\\
    \caption{SLL versus sum SE Pareto space of (a)~DTPA with $2\times10^6$ points, (b)~TTPA with $8\times10^5$ points, (c)~$\rho~=~0.5$ thinned array ($S~=~40$) with $10^5$ points, (d)~$\rho~=~0.25$ thinned array ($S~=~20$) with $8\times10^5$ points, respectively.}
    \label{fig:Pareto}
\end{figure}

Fig.~\ref{fig:Pareto} illustrates the SLL versus the sum SE Pareto space for the following FD configurations:~(a)~DTPA with $2\times10^6$ points, (b)~TTPA with $8\times10^5$ points, (c)~50\% thinned array with $10^5$ points and (d)~25\% thinned array with $8\times10^5$ points. The color coding represents the distribution of samples along the sum SE and SLL axes. It is important to note that in Fig.~\ref{fig:Pareto}(a), the DTPA solution space exhibits discrete behavior along the SE axis due to having only vertical and horizontal unit elements, while in Fig.~\ref{fig:Pareto}(b), the TTPA solution space follows a Gaussian distribution over both axes since the phase center of the unit elements can be better randomized. The optimal configurations for different $\beta$ values are indicated with a square marker, showing how the trade-off between sum SE and SLL performance forms a Pareto front. The optimized DTPA configurations found by GA and Algorithm\--X techniques are indicated with cross and square markers in Fig.~\ref{fig:Pattern}(a), respectively. Similarly, the optimized TTPA configurations found by GA and Algorithm\--X techniques are indicated with cross and square markers in Fig.~\ref{fig:Pareto}(b), respectively. We observe that Algorithm\--X can find better solutions compared to GA based search at the expense of increased complexity. 

Next, the focus is placed on comparing hybrid and digital architectures in combination with irregular arrays. The same optimization process has been performed for each architectural option, and the resulting normalized power patterns have been compared. For illustration, a single channel realization has been selected from the 25 realizations used during optimization. Specifically, the selected channel realization involves two receivers located at angular coordinates 
$\left(u_1,v_1\right)~=~\left(0.25,0.25\right)$ and $\left(u_2,v_2\right)~=~\left(-0.25,-0.75\right)$, respectively. The optimal array configuration for each architecture is plotted, with $\beta~=~0$ is considered in the objective function~\eqref{eq:OptProb}. Algorithm\--X based optimization results are reported for clustered arrays while the method in~\cite{Zhang2023} is considered for thinned array optimization.

\begin{figure}[]
    \centering
    \includegraphics[width=0.32\linewidth]{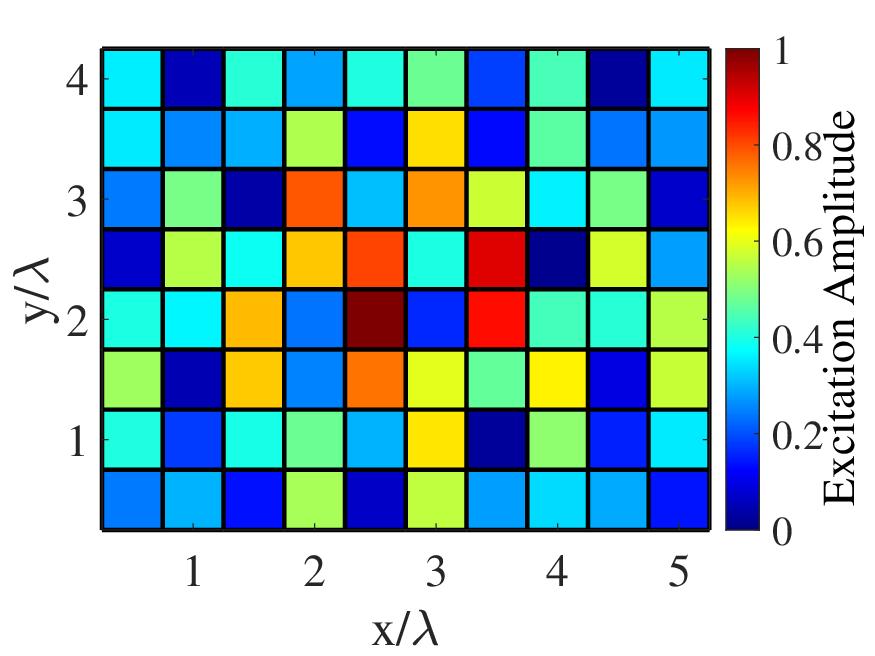}
    \includegraphics[width=0.32\linewidth]{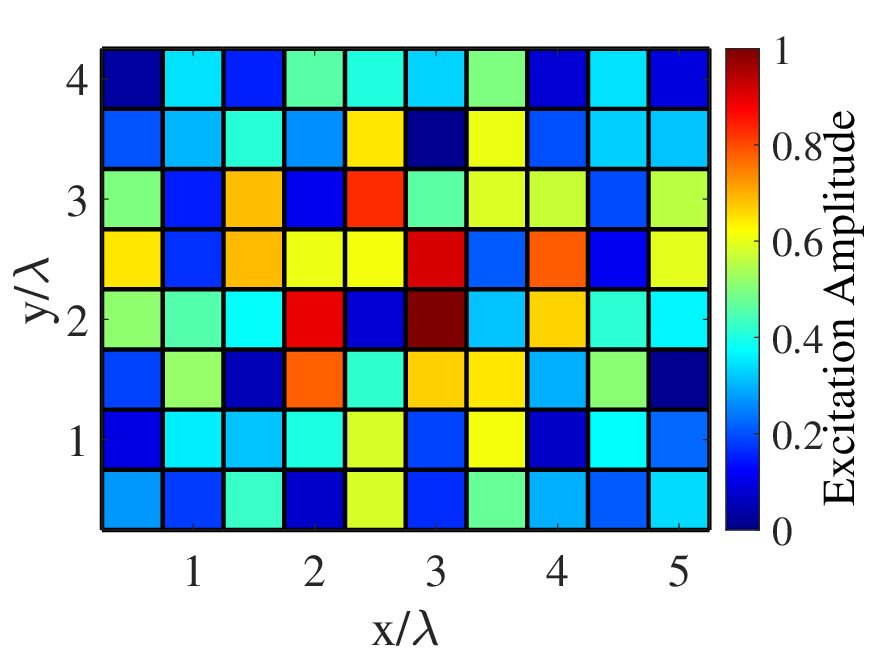}
    \includegraphics[width=0.32\linewidth]{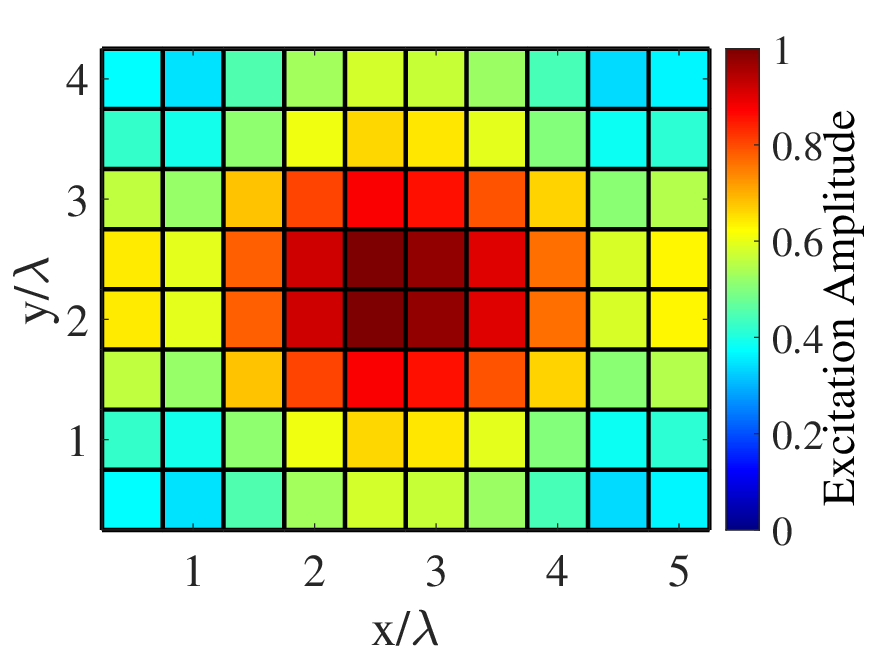}\\   
    \small(a)\hspace{70pt}(b)\hspace{70pt}(c)\\
    \includegraphics[width=0.32\linewidth]{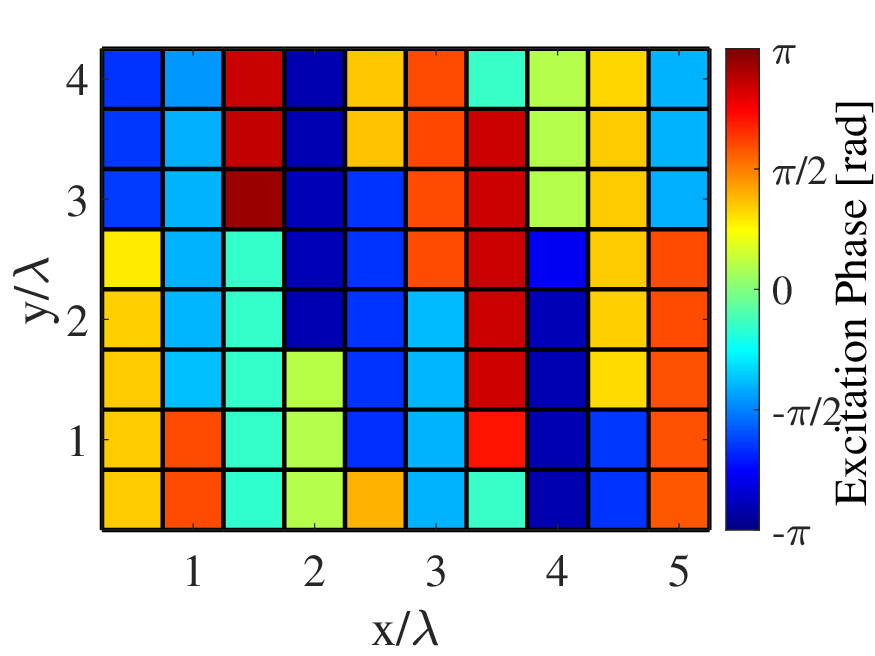}
    \includegraphics[width=0.32\linewidth]{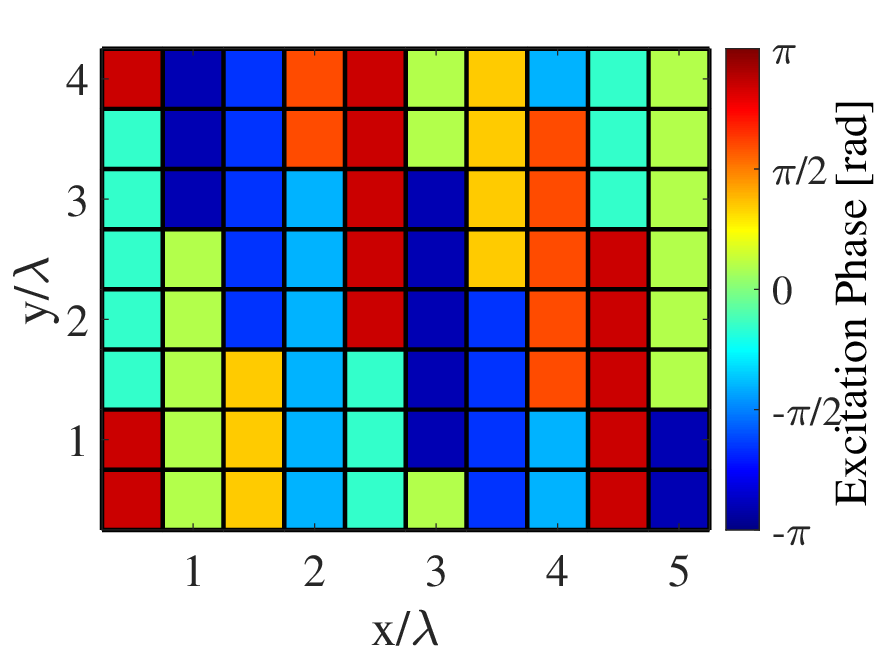}
    \includegraphics[width=0.32\linewidth]{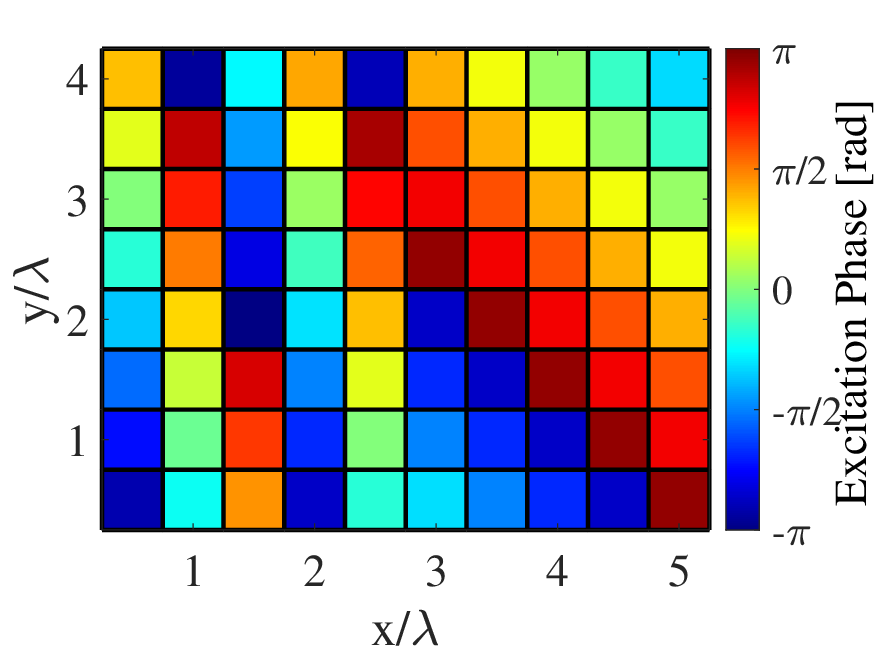}\\
    \small(d)\hspace{70pt}(e)\hspace{70pt}(f)\\
    \includegraphics[width=0.32\linewidth]{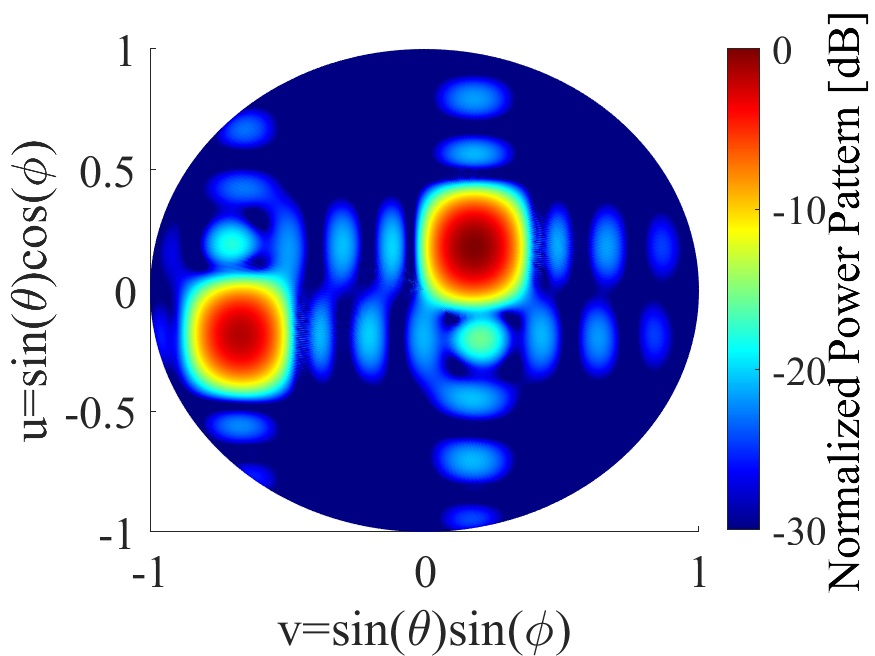}
    \includegraphics[width=0.32\linewidth]{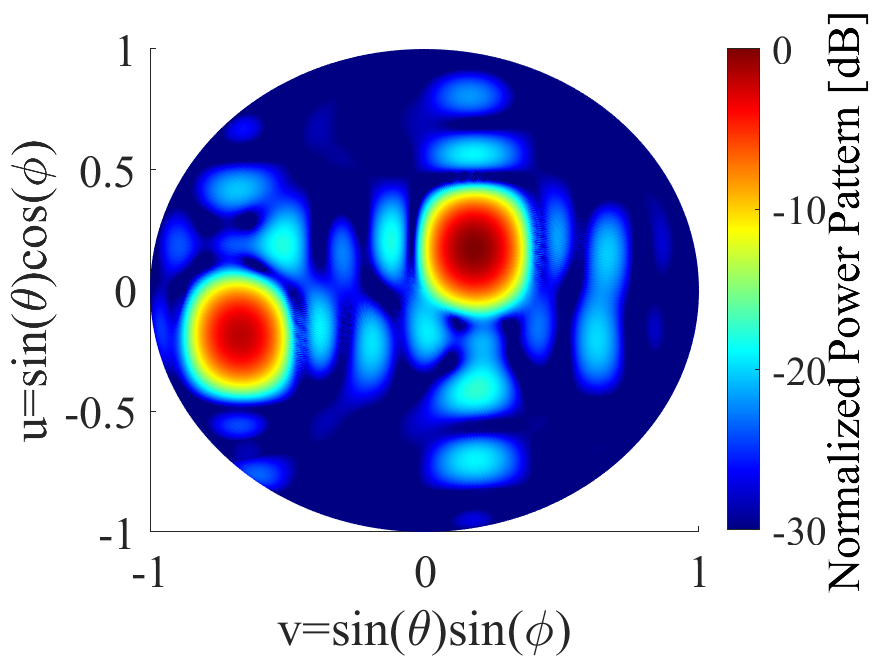}
    \includegraphics[width=0.32\linewidth]{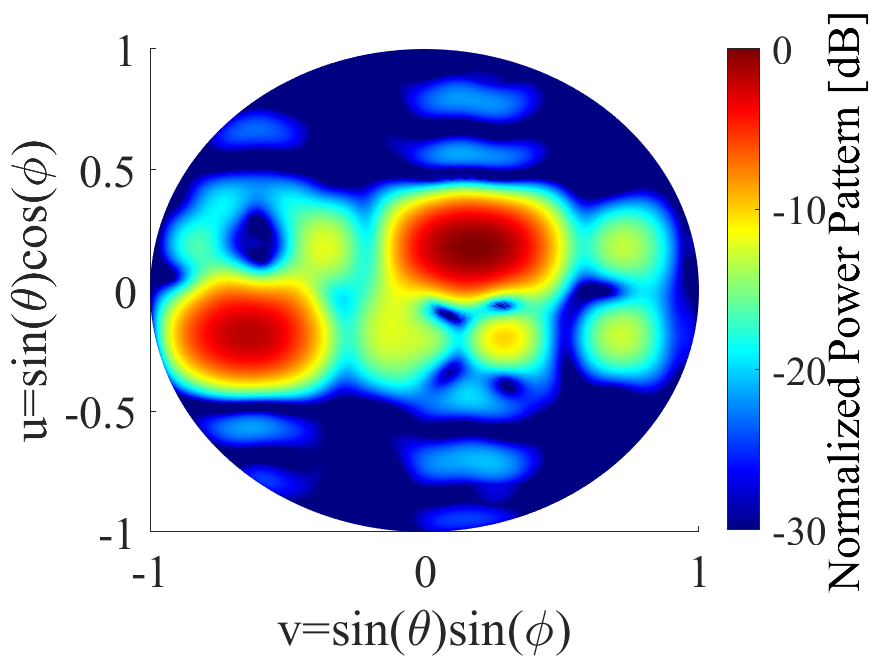}\\
    \small(g)\hspace{70pt}(h)\hspace{70pt}(i)\\
    \caption{Optimized $(N~\times~M)~=~8~\times~10$ FPRA (a)~Excitation amplitude, (d)~excitation phase, and (g)~normalized power pattern of FD architecture, (b)~excitation amplitude, (e)~excitation phase, (h)~normalized power pattern of HFC architecture, (c)~excitation amplitude, (f)~excitation phase, and (i)~normalized power pattern of HPC architecture, respectively.}
    \label{fig:FPRAres}
\end{figure}

Fig.~\ref{fig:FPRAres} depicts the detailed array excitation and resulting normalized power pattern of the optimized $(N~\times~M)~=~8~\times~10$ FPRA architectures. Figs.~\ref{fig:FPRAres}(a), (d), and (g) depict the excitation amplitude, phase, and normalized power pattern for the FPRA with FD architecture, respectively. Figs.~\ref{fig:FPRAres}(b), (e), and (h) depict the excitation amplitude, phase, and normalized power pattern for the FPRA with HFC architecture, respectively. Figs.~\ref{fig:FPRAres}(c), (f), and (i) depict the excitation amplitude, phase, and normalized power pattern for the FPRA with HPC architecture, respectively. FD and HFC architectures exhibit two clear beams toward two receivers while maintaining low SLLs, while the HPC architecture has wider beams and increased SLLs due to limited array excitation control. However, realizing these array architectures is challenging considering the beyond 100~GHz implementation constraints. 

\begin{figure}[]
\centering
    \includegraphics[width=0.32\linewidth]{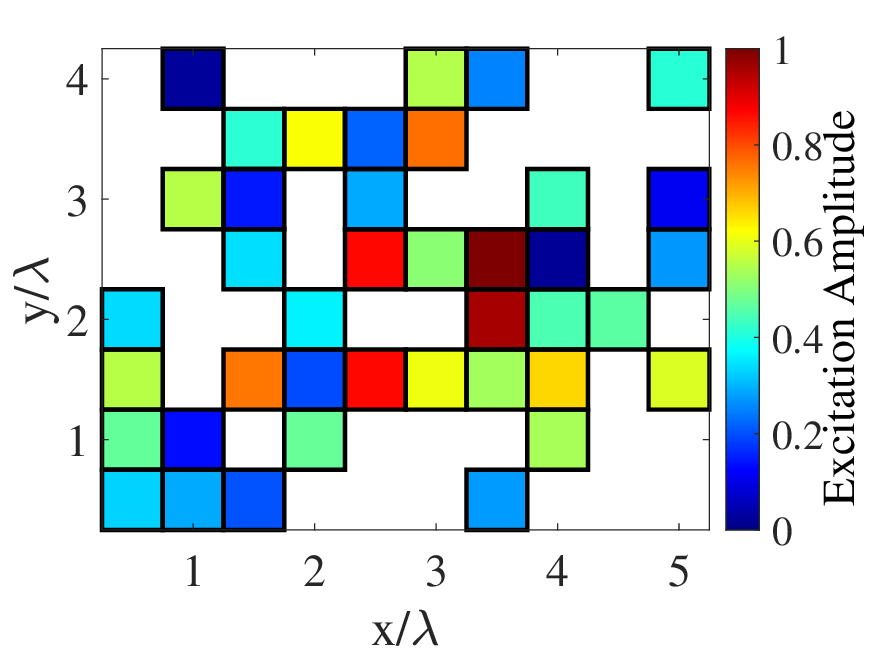}
    \includegraphics[width=0.32\linewidth]{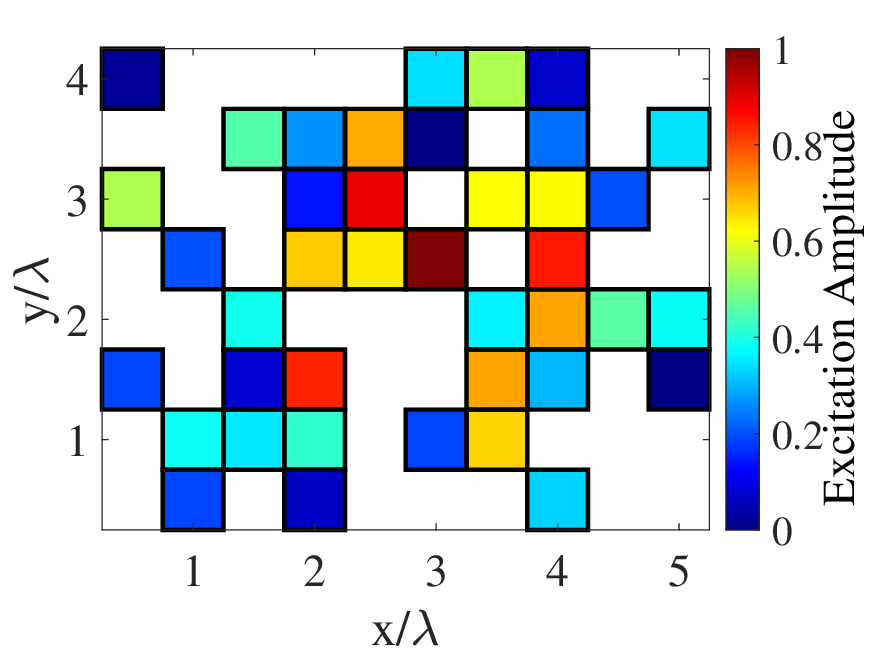}
    \includegraphics[width=0.32\linewidth]{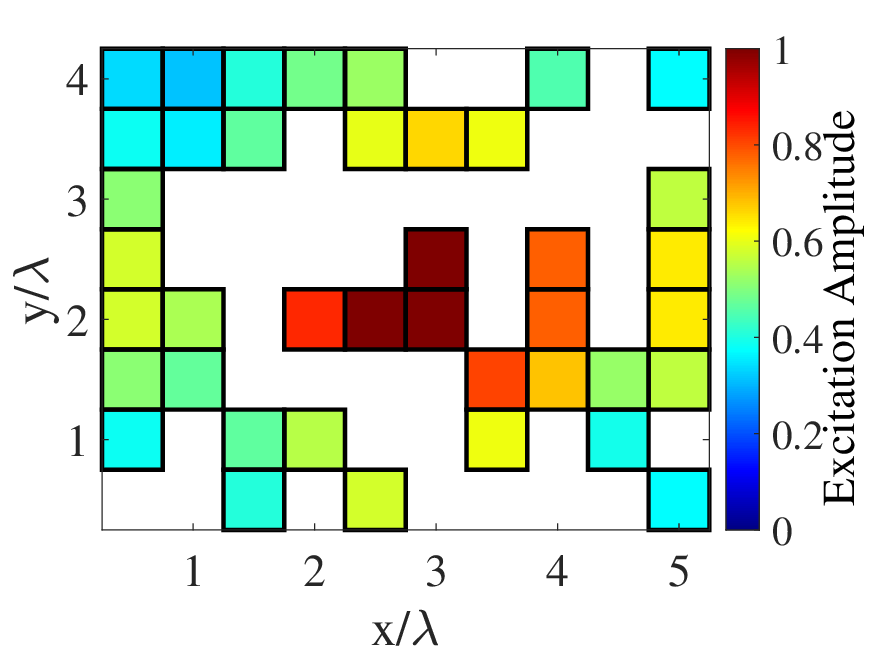}\\
     \small(a)\hspace{70pt}(b)\hspace{70pt}(c)\\
    \includegraphics[width=0.32\linewidth]{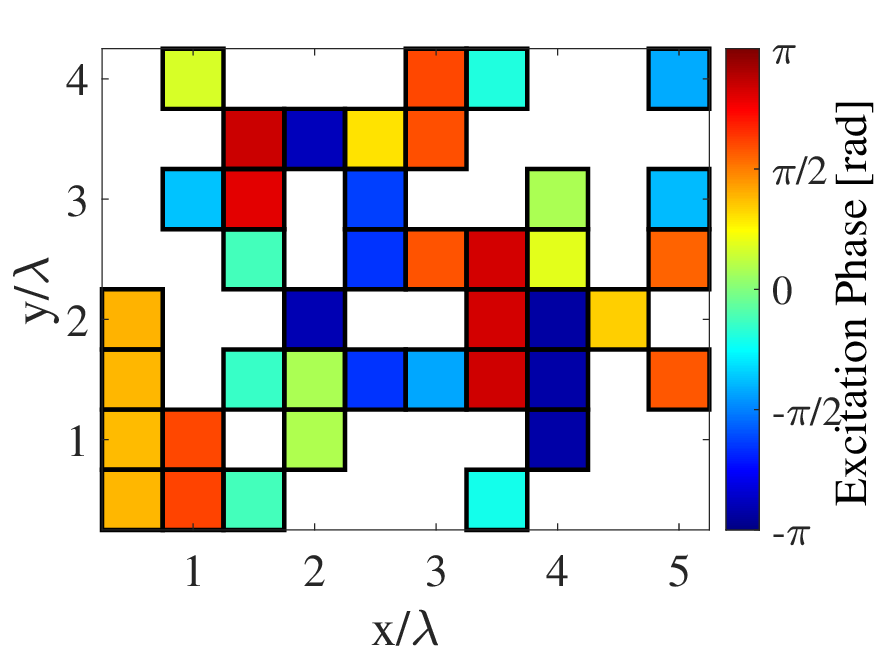}
    \includegraphics[width=0.32\linewidth]{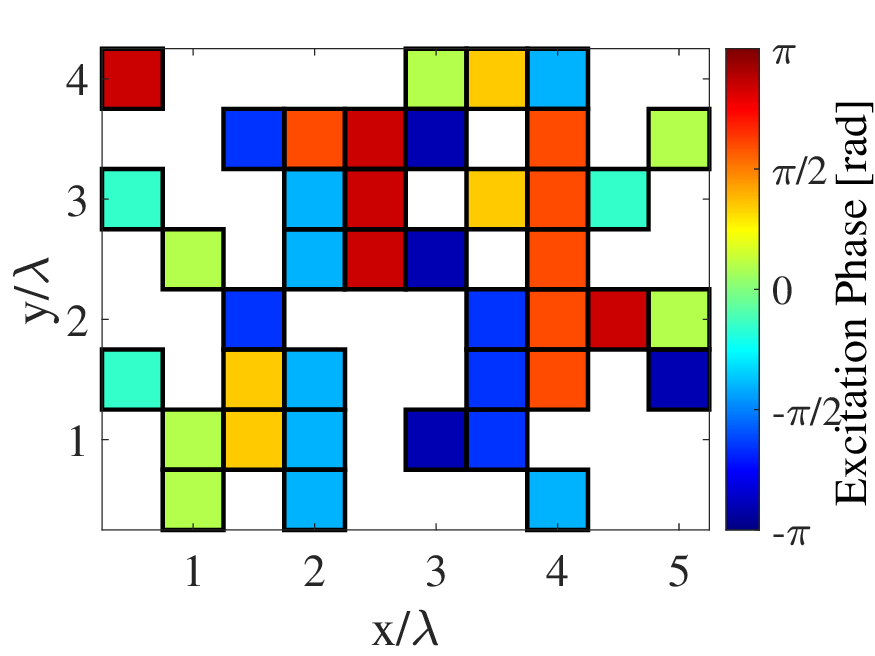}
    \includegraphics[width=0.32\linewidth]{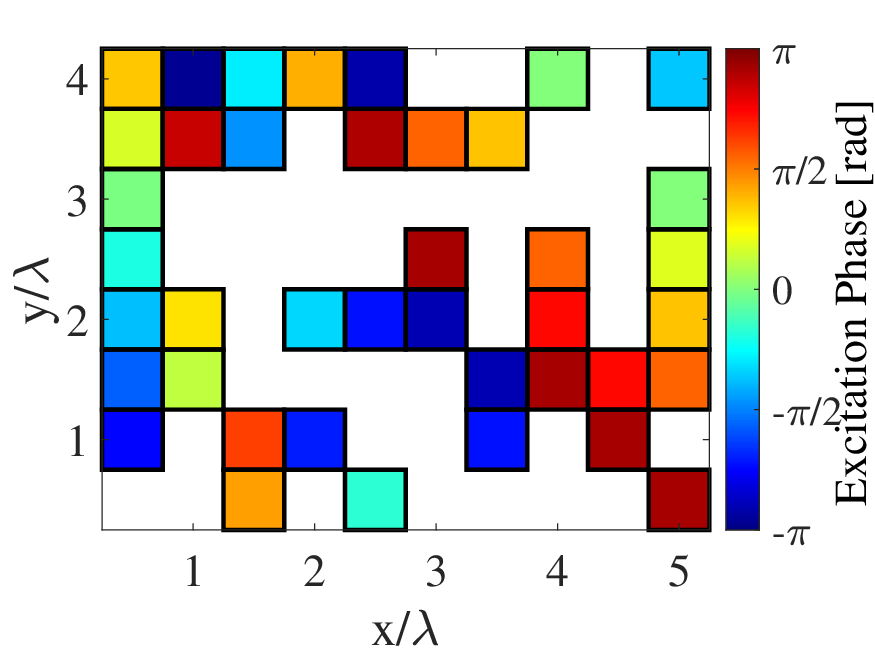}\\
    \small(d)\hspace{70pt}(e)\hspace{70pt}(f)\\
    \includegraphics[width=0.32\linewidth]{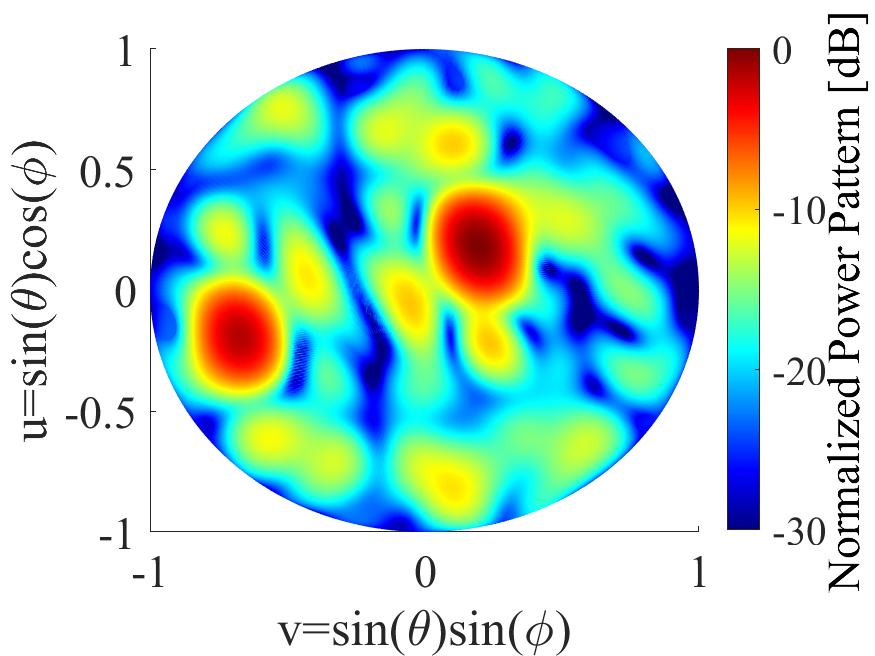}
    \includegraphics[width=0.32\linewidth]{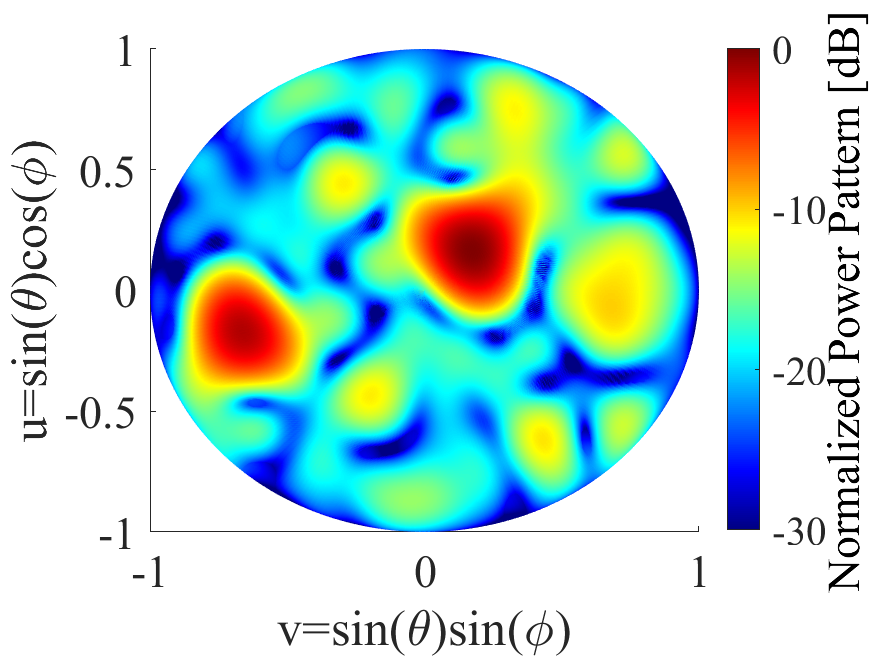}
    \includegraphics[width=0.32\linewidth]{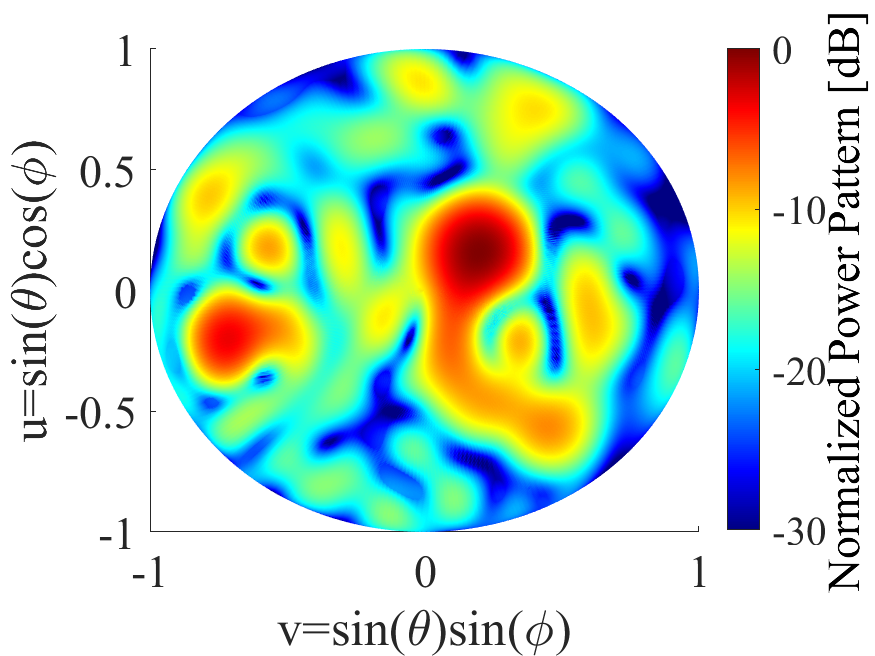}\\
    \small(g)\hspace{70pt}(h)\hspace{70pt}(i)\\
    \caption{Optimized $(N~\times~M)~=~8~\times~10$ thinned array with $S~=~40$, (a)~Excitation amplitude, (d)~excitation phase, and (g)~normalized power pattern of FD architecture, (b)~excitation amplitude, (e)~excitation phase, (h)~normalized power pattern of HFC architecture, (c)~excitation amplitude, (f)~excitation phase, and (i)~normalized power pattern of HPC architecture, respectively.}
    \label{fig:ThinS40}
\end{figure}

Fig.~\ref{fig:ThinS40} depicts the detailed array excitation and resulting normalized power pattern of optimized $(N~\times~M)~=~8~\times~10$, $\rho~=~0.5$ thinned array $(S~=~40)$ architectures. Figs.~\ref{fig:ThinS40}(a), (d), and (g) depict the excitation amplitude, phase, and normalized power pattern for the thinned array with FD architecture, respectively. Figs.~\ref{fig:ThinS40}(b), (e), and (h) depict the excitation amplitude, phase, and normalized power pattern for the thinned array with HFC architecture, respectively. Figs.~\ref{fig:ThinS40}(c), (f), and (i) depict the excitation amplitude, phase, and normalized power pattern for the thinned array with HPC architecture, respectively. In all cases, one can identify two beams towards two receivers where the optimal array configuration changes with the architectural choice. In particular, HPC architecture exhibits higher SLLs due to reduced control on the array excitation, but the size of the main beams does not enlarge in contrast to the FPRA array with HPC architecture.

\begin{figure}[]
\centering
    \includegraphics[width=0.32\linewidth]{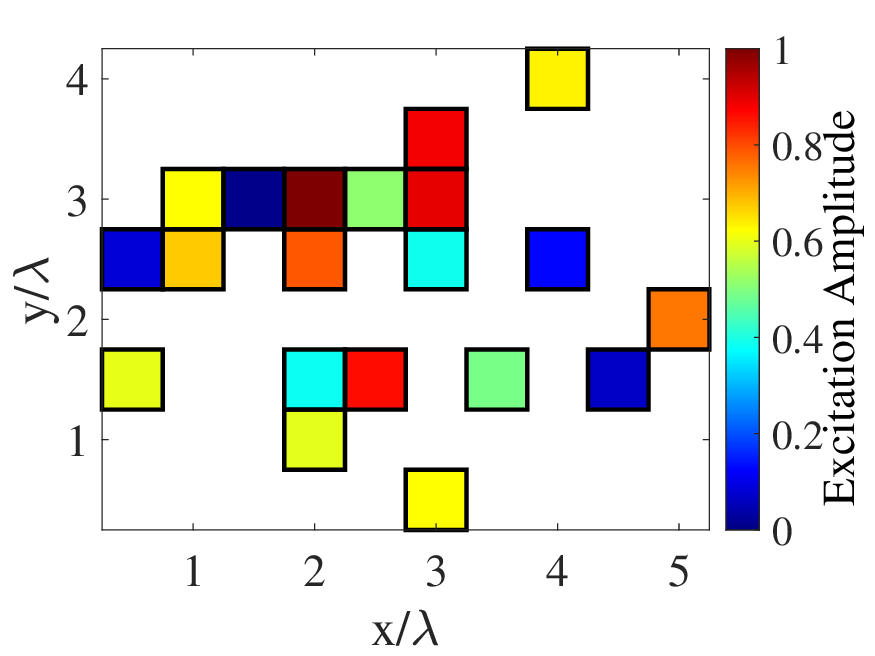}
    \includegraphics[width=0.32\linewidth]{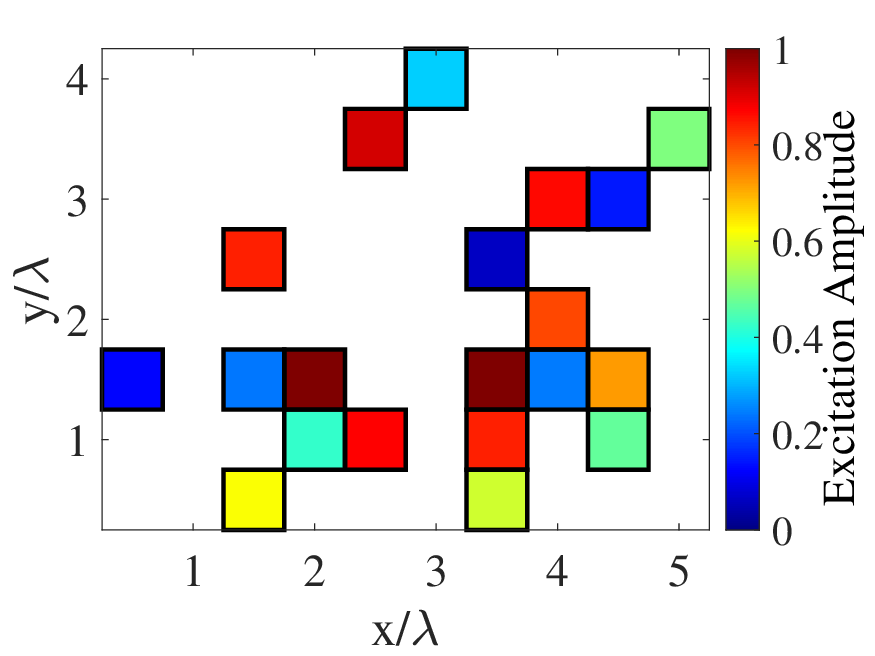}
    \includegraphics[width=0.32\linewidth]{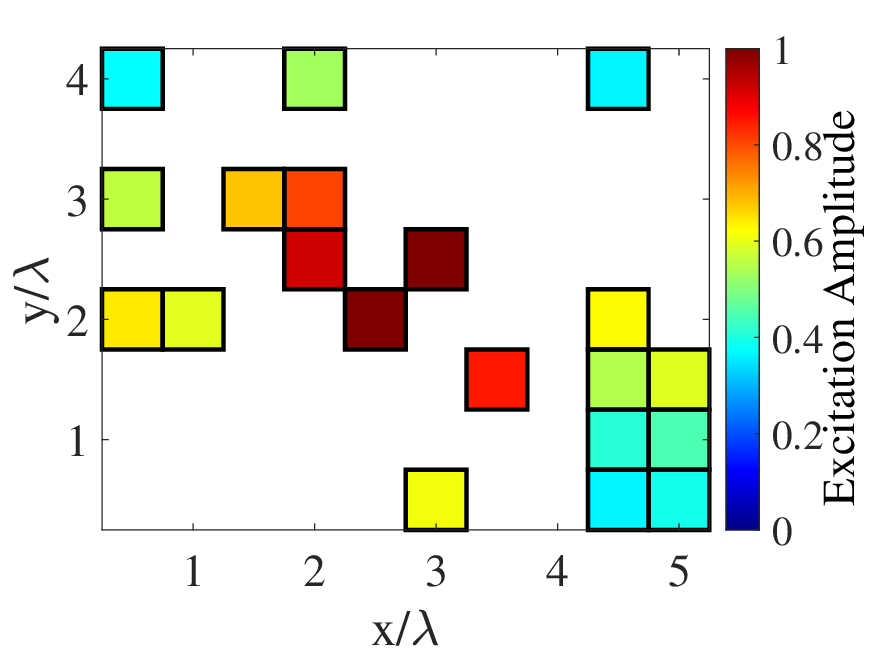}\\
     \small(a)\hspace{70pt}(b)\hspace{70pt}(c)\\
    \includegraphics[width=0.32\linewidth]{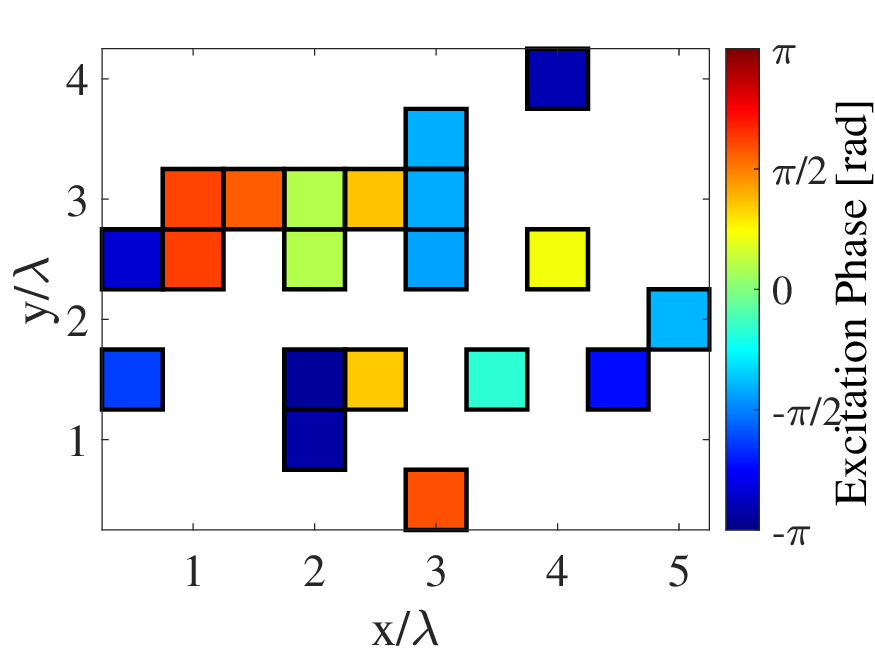}
    \includegraphics[width=0.32\linewidth]{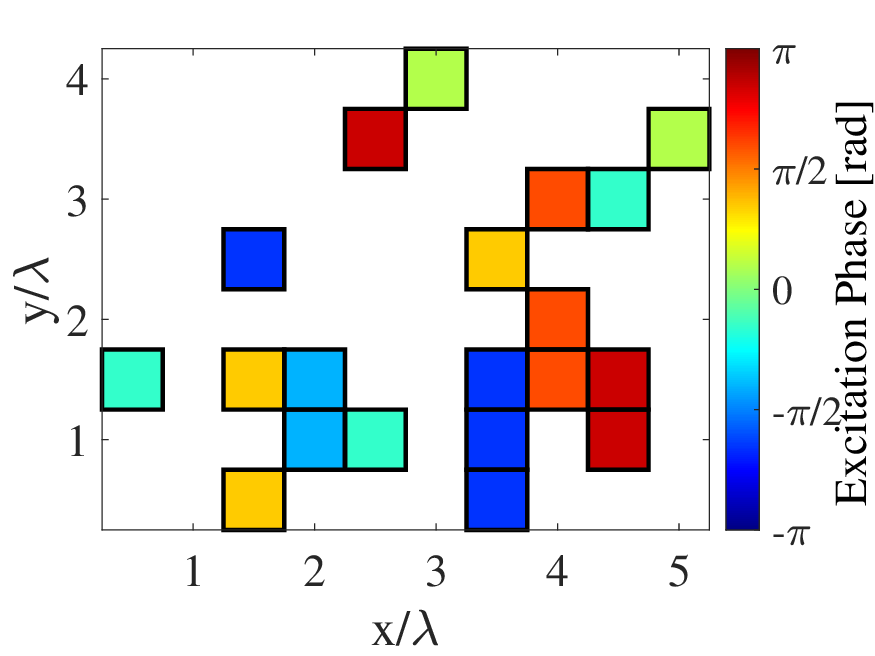}
    \includegraphics[width=0.32\linewidth]{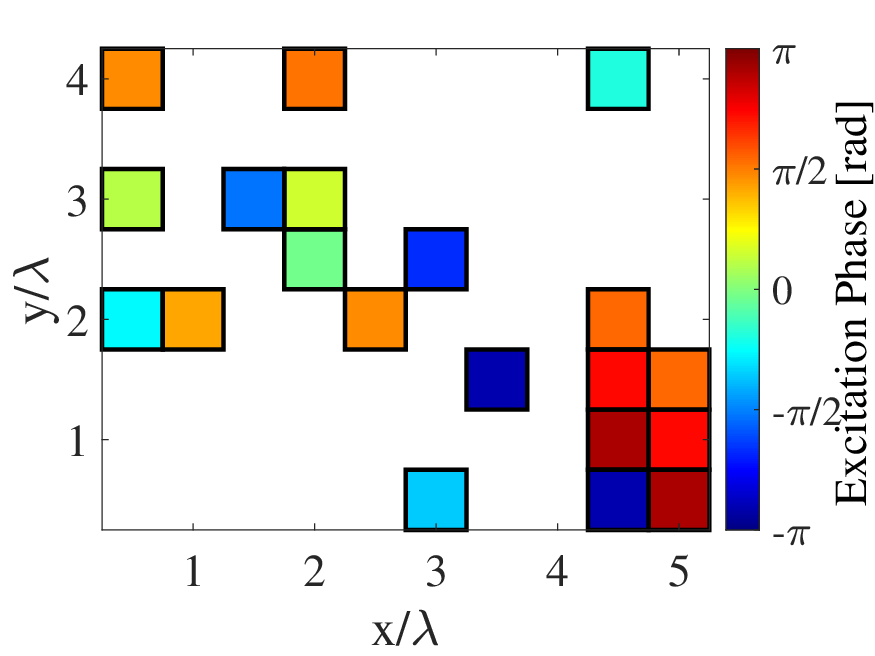}\\
    \small(d)\hspace{70pt}(e)\hspace{70pt}(f)\\
    \includegraphics[width=0.32\linewidth]{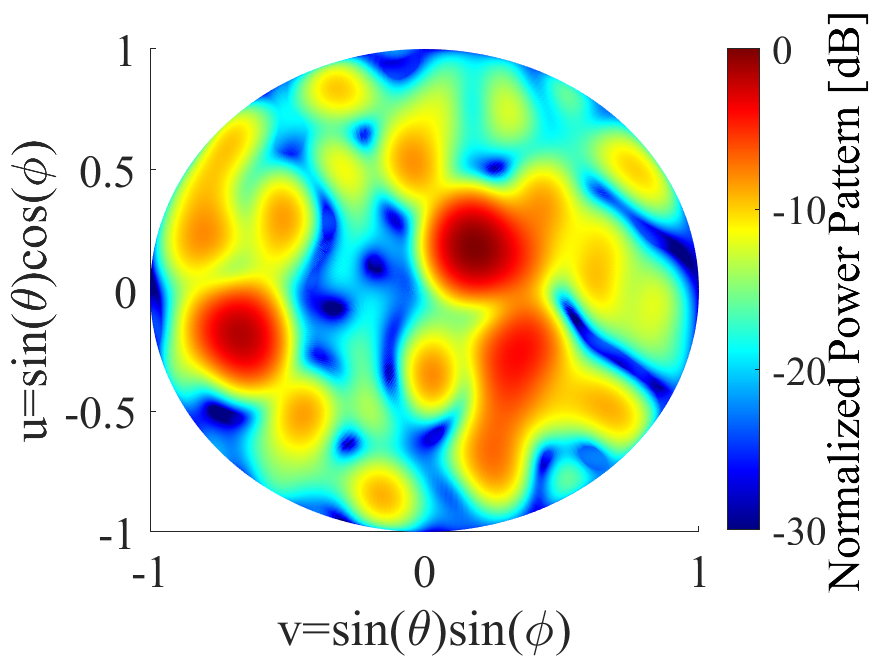}
    \includegraphics[width=0.32\linewidth]{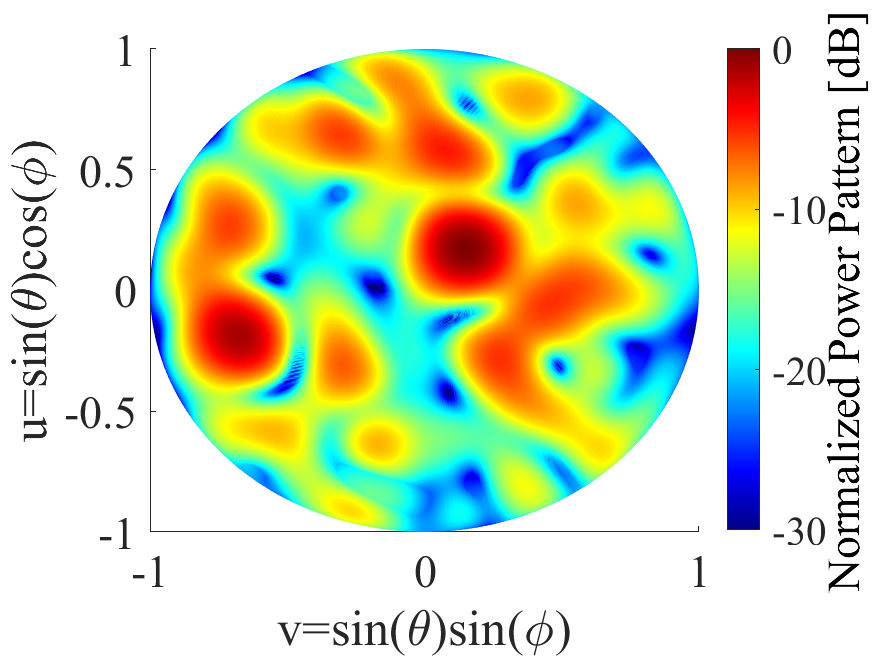}
    \includegraphics[width=0.32\linewidth]{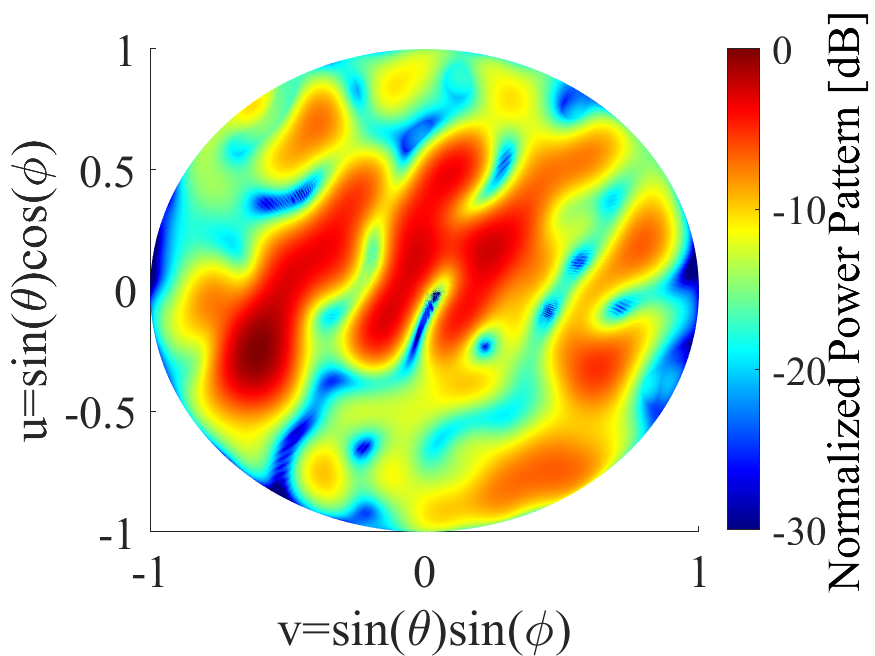}\\
    \small(g)\hspace{70pt}(h)\hspace{70pt}(i)\\
    \caption{Optimized $(N~\times~M)~=~8~\times~10$ thinned array with $S~=~20$: (a)~Excitation amplitude, (d)~excitation phase, and (g)~normalized power pattern of FD architecture, (b)~excitation amplitude, (e)~excitation phase, (h)~normalized power pattern of HFC architecture, (c)~excitation amplitude, (f)~excitation phase, and (i)~normalized power pattern of HPC architecture, respectively.}
    \label{fig:ThinS20}
\end{figure}

Fig.~\ref{fig:ThinS20} depicts the detailed array excitation and resulting normalized power pattern of optimized $(N~\times~M)~=~8~\times~10$, $\rho~=~0.25$ thinned array $(S~=~20)$ architectures. Figs.~\ref{fig:ThinS20}(a), (d), and (g) depict the excitation amplitude, phase, and normalized power pattern for the thinned array with FD architecture, respectively. Figs.~\ref{fig:ThinS20}(b), (e), and (h) depict the excitation amplitude, phase, and normalized power pattern for the thinned array with HFC architecture, respectively. Figs.~\ref{fig:ThinS20}(c), (f), and (i) depict the excitation amplitude, phase, and normalized power pattern for the thinned array with HPC architecture, respectively. Thinned arrays with FD and HFC architecture exhibit two beams towards two receivers at the expense of high SLLs. Furthermore, the thinned array with HPC cannot accommodate two clear beams due to a low thinning ratio combined with the HPC architecture.

\begin{figure}[]
    \centering
    \includegraphics[width=0.32\linewidth]{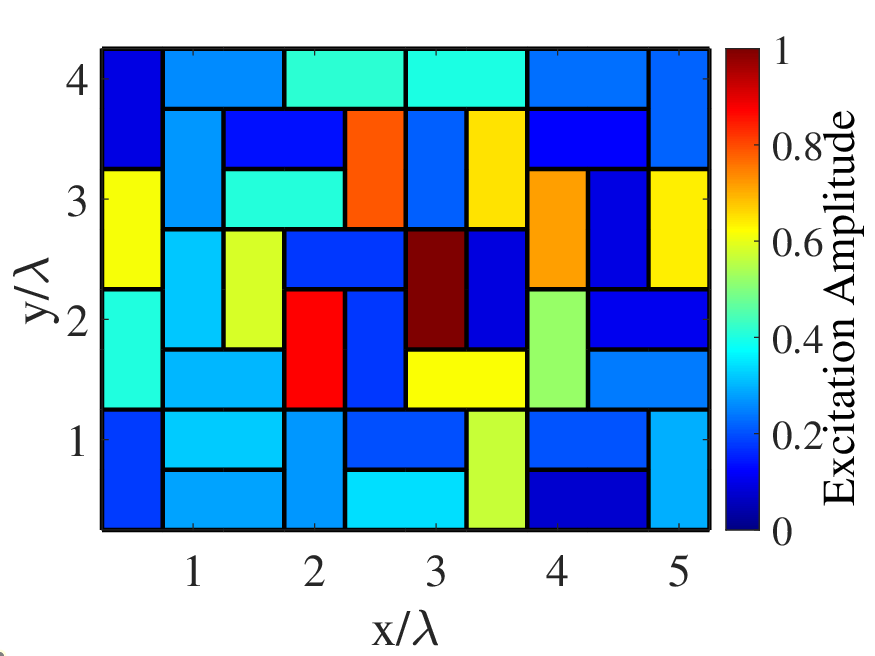}
    \includegraphics[width=0.32\linewidth]{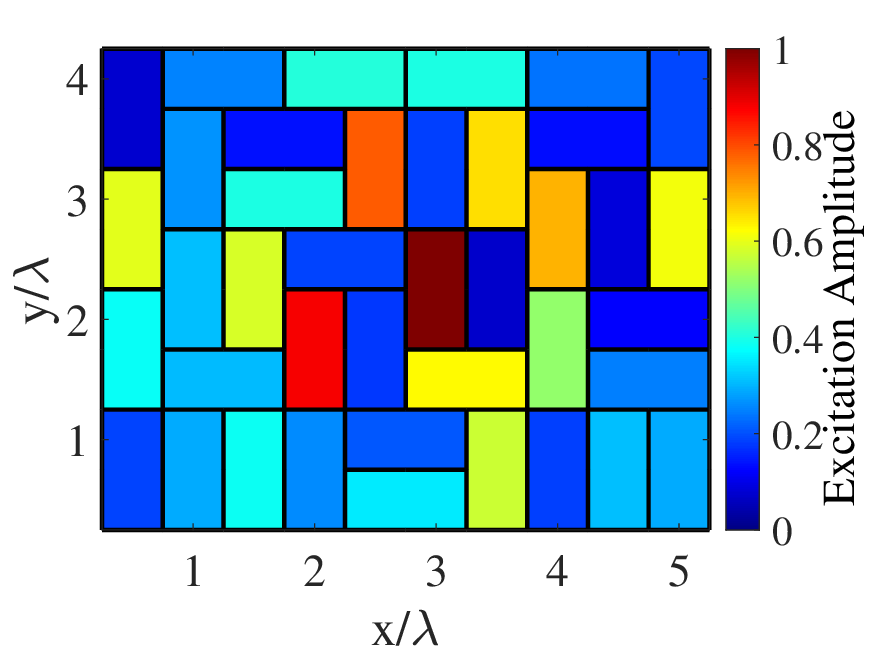}
    \includegraphics[width=0.32\linewidth]{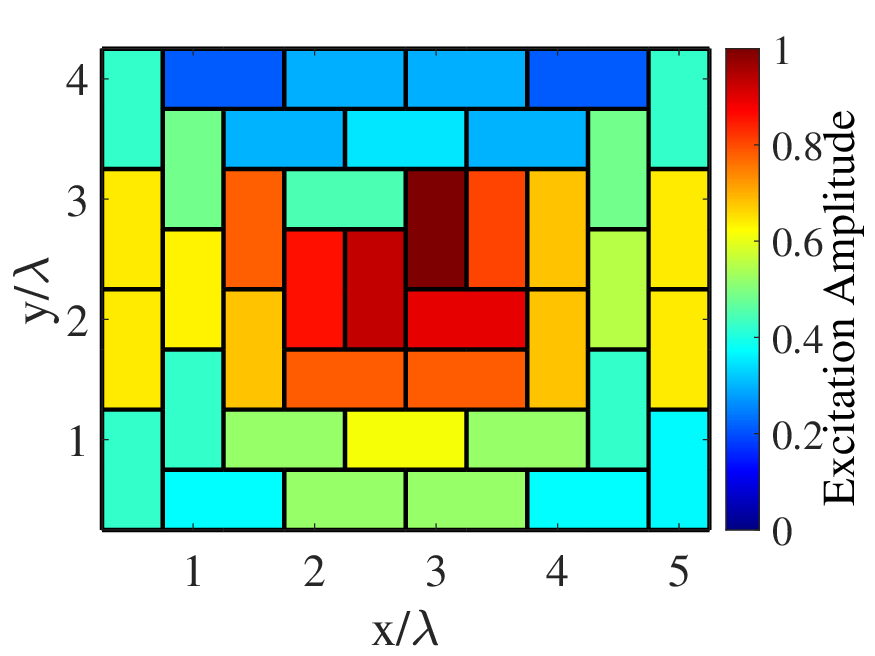}\\
    \small(a)\hspace{70pt}(b)\hspace{70pt}(c)\\
    \includegraphics[width=0.32\linewidth]{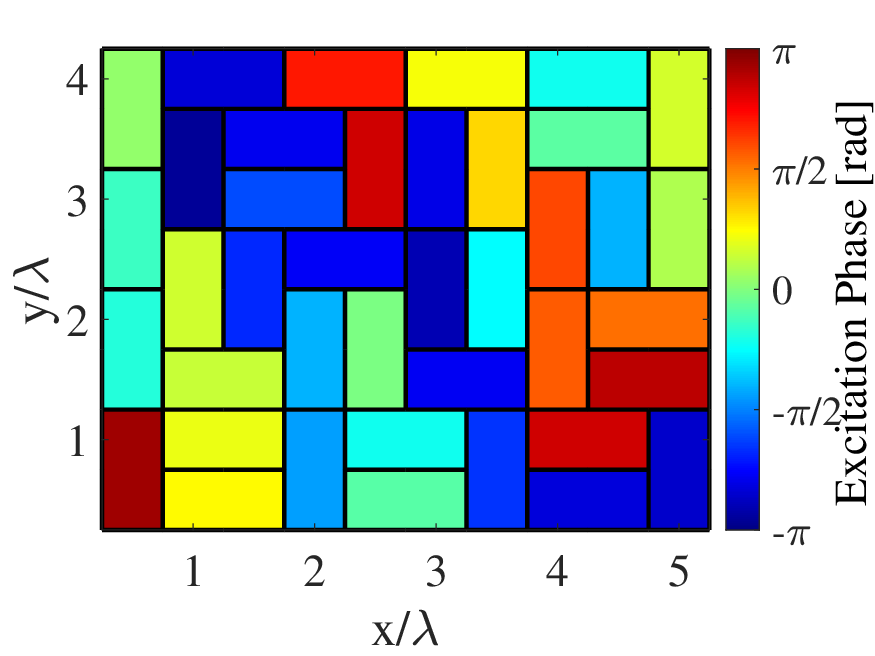}
    \includegraphics[width=0.32\linewidth]{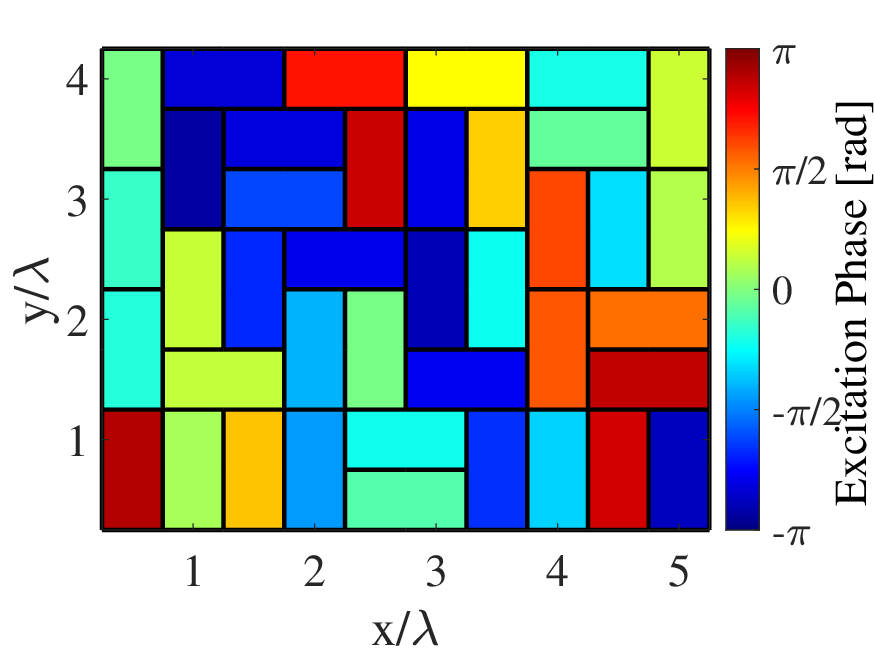}
    \includegraphics[width=0.32\linewidth]{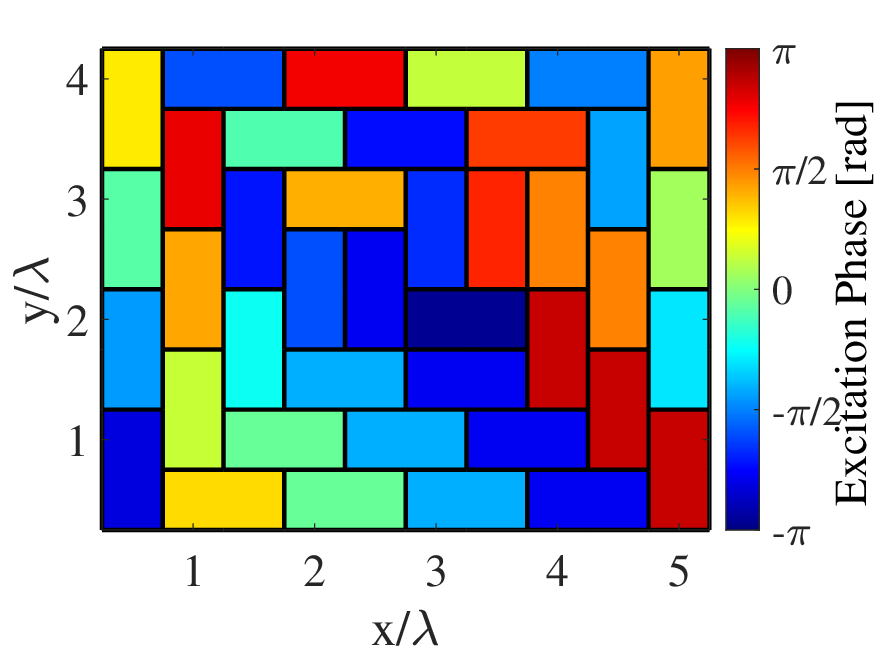}\\
    \small(d)\hspace{70pt}(e)\hspace{70pt}(f)\\
    \includegraphics[width=0.32\linewidth]{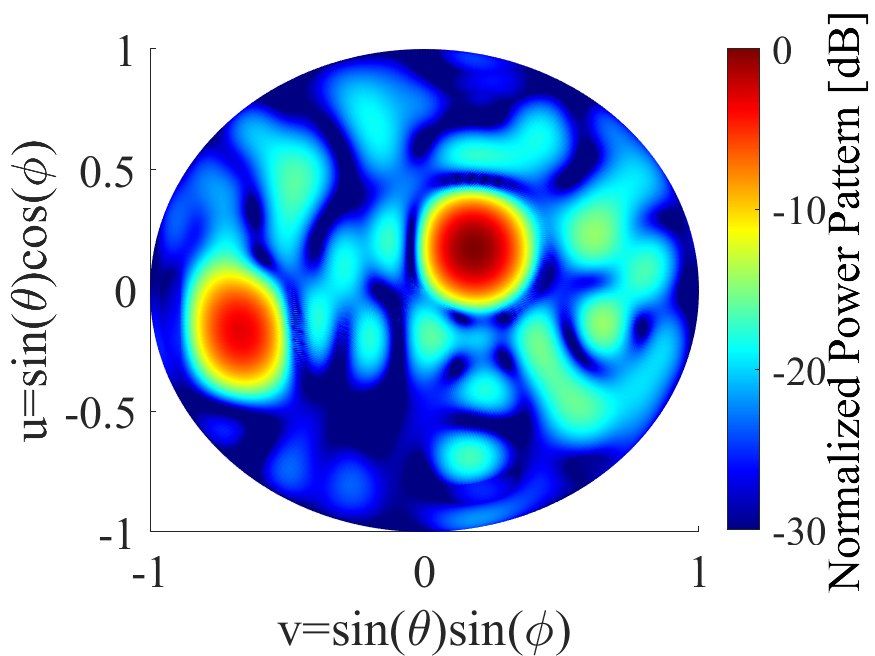}
    \includegraphics[width=0.32\linewidth]{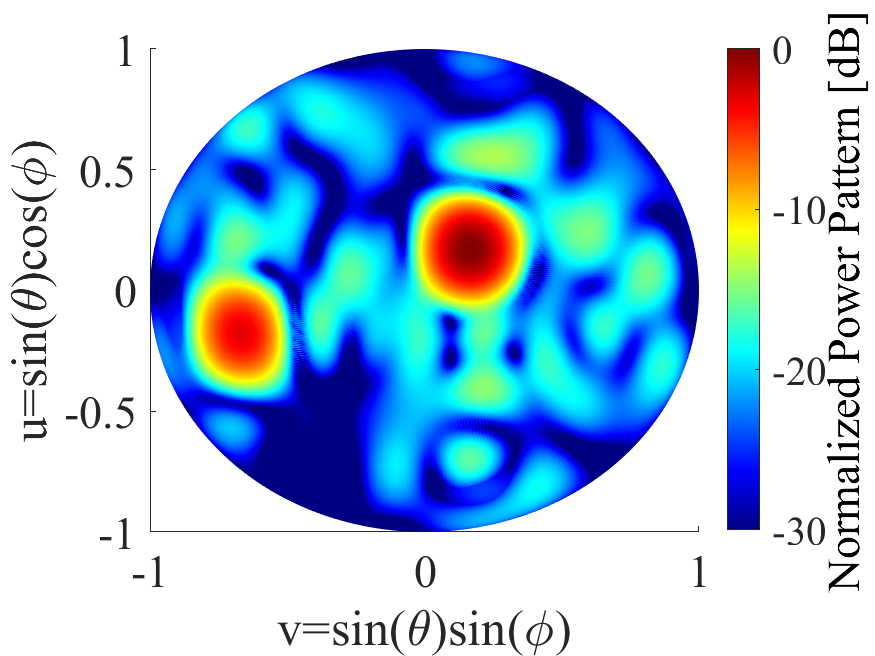}
    \includegraphics[width=0.32\linewidth]{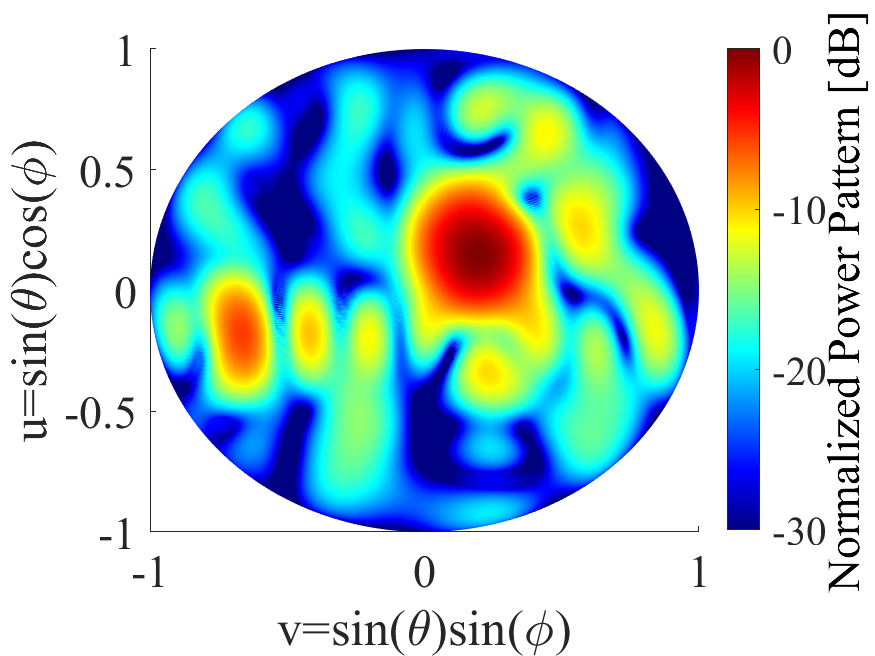}\\
    \small(g)\hspace{70pt}(h)\hspace{70pt}(i)\\
    \caption{Optimized $(N~\times~M)~=~8~\times~10$ DTPA, (a) Excitation amplitude, (d) excitation phase, and (g) normalized power pattern of FD architecture, (b) excitation amplitude, (e) excitation phase, (h) normalized power pattern of HFC architecture, (c) excitation amplitude, (f) excitation phase, and (i) normalized power pattern of HPC architecture, respectively.}
    \label{fig:DTPA}
\end{figure}

Fig.~\ref{fig:DTPA} depicts the detailed array excitation and normalized power pattern of optimized $(N~\times~M)~=~8~\times~10$ DTPA architectures. Figs.~\ref{fig:DTPA}(a), (d), and (g) depict the excitation amplitude, phase, and normalized power pattern for the DTPA with FD architecture, respectively. Figs.~\ref{fig:DTPA}(b), (e), and (h) depict the excitation amplitude, phase, and normalized power pattern for the DTPA with HFC architecture, respectively. Figs.~\ref{fig:DTPA}(c), (f), and (i) depict the excitation amplitude, phase, and normalized power pattern for the DTPA with HPC architecture, respectively. All DTPA architectures maintain low SLLs compared to the thinned array with $\rho~=~0.5$. DTPA with FD and HFC architecture maintains two clear beams towards two receivers, while the control on the second beam in the HPC architecture is significantly reduced.

\begin{figure}[t!]
    \centering
    \includegraphics[width=0.32\linewidth]{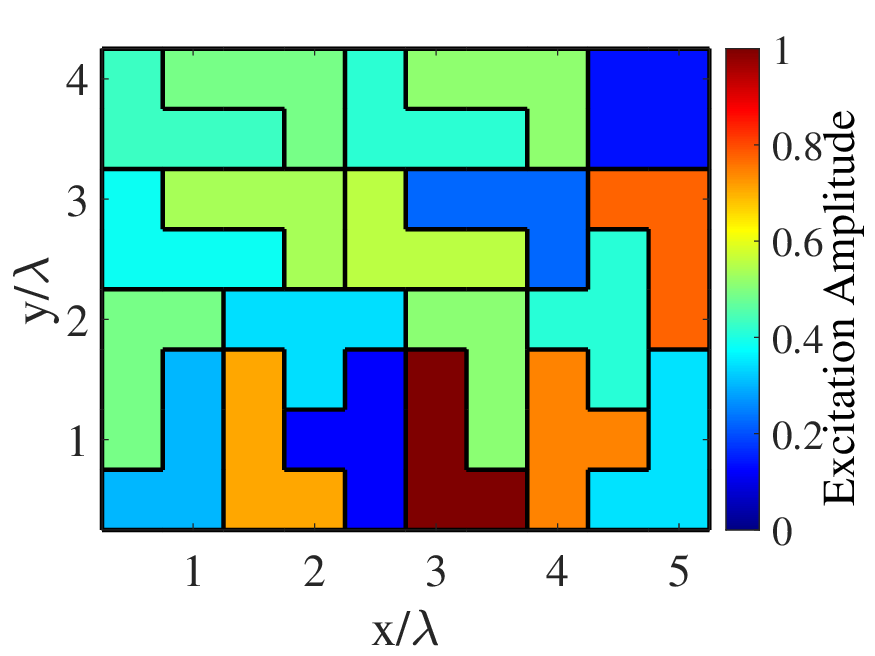}
    \includegraphics[width=0.32\linewidth]{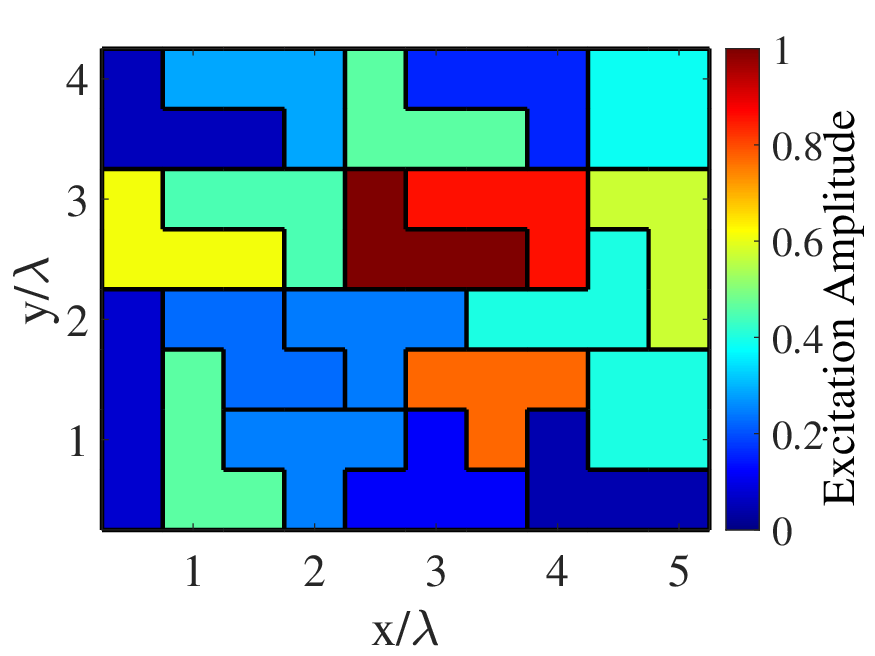}
    \includegraphics[width=0.32\linewidth]{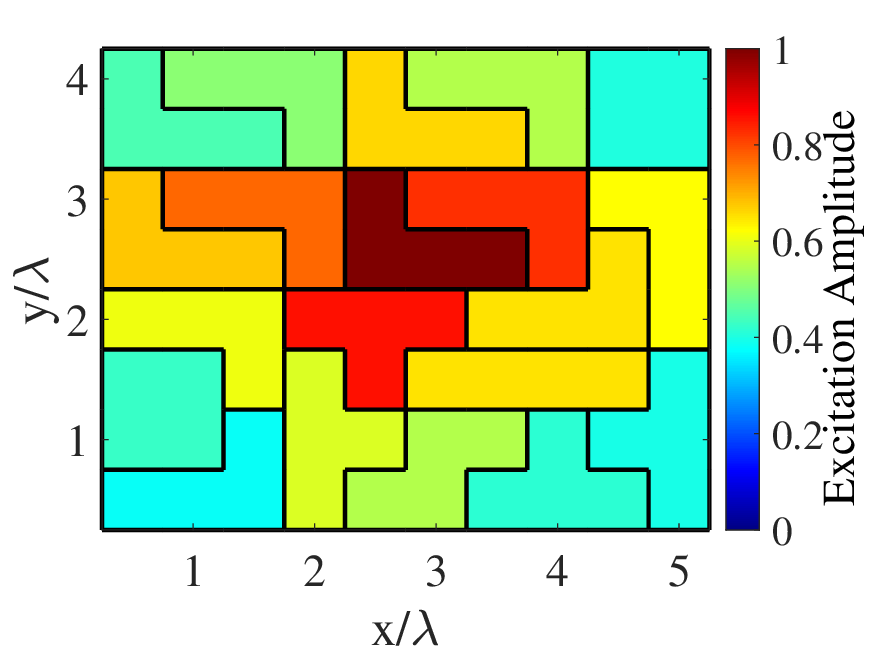}\\
    
    \small(a)\hspace{70pt}(b)\hspace{70pt}(c)\\
    \includegraphics[width=0.32\linewidth]{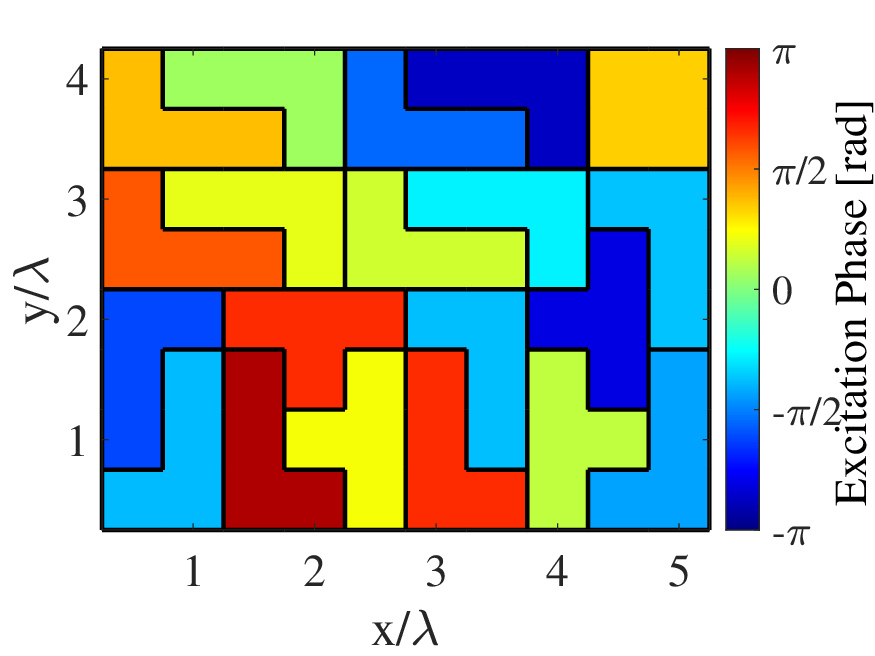}
    \includegraphics[width=0.32\linewidth]{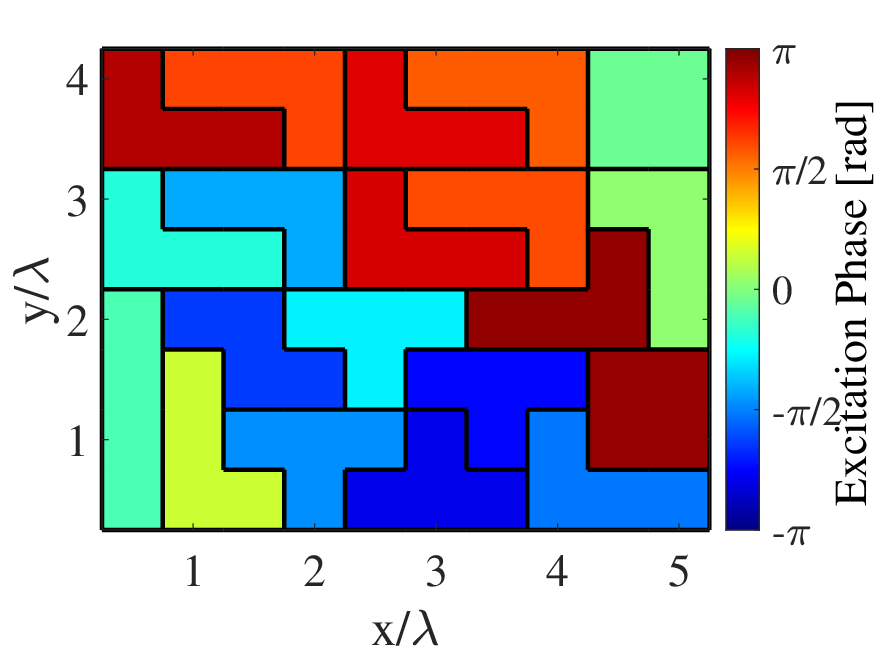}
     \includegraphics[width=0.32\linewidth]{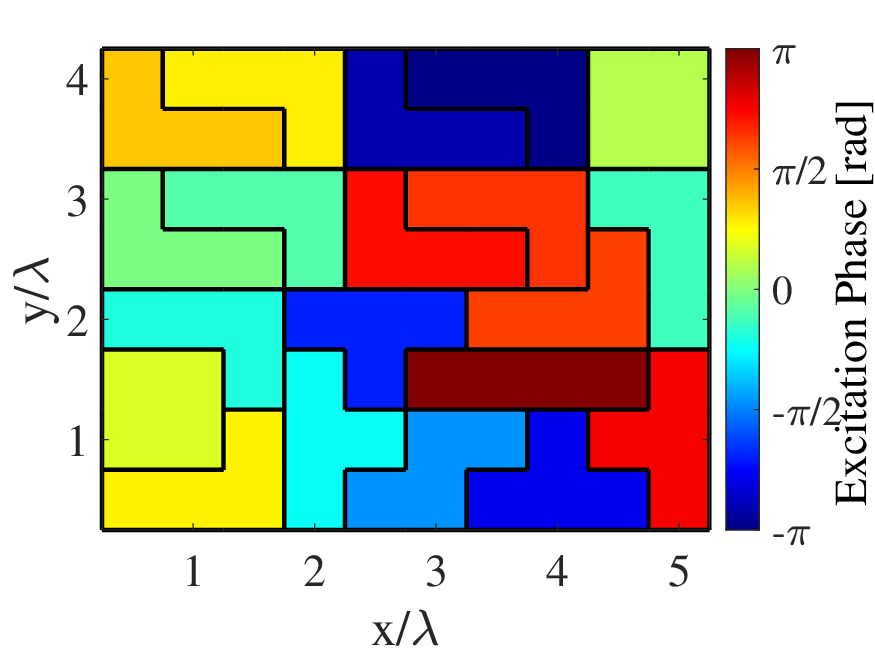}\\
    \small(d)\hspace{70pt}(e)\hspace{70pt}(f)\\
     \includegraphics[width=0.32\linewidth]{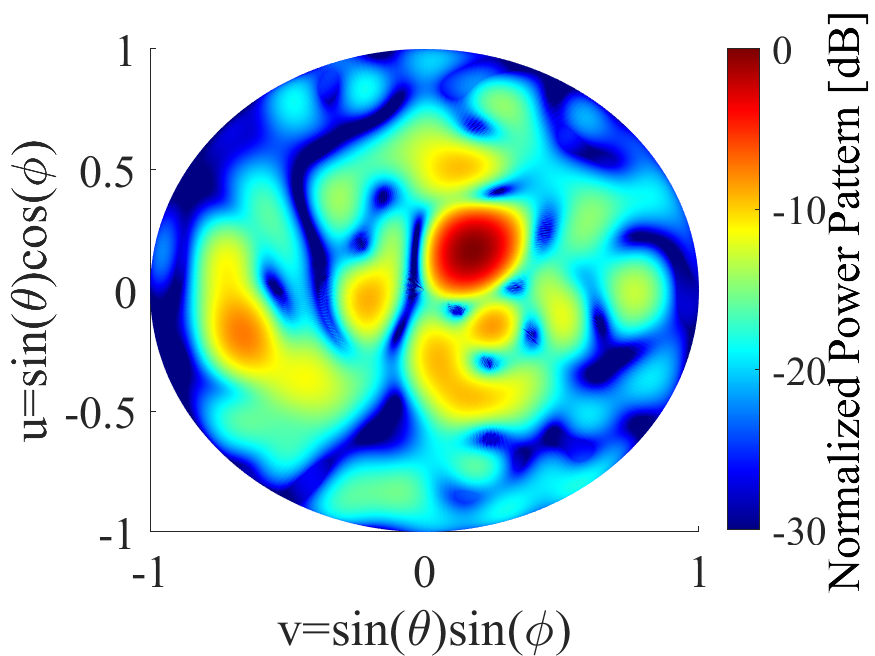}
   \includegraphics[width=0.32\linewidth]{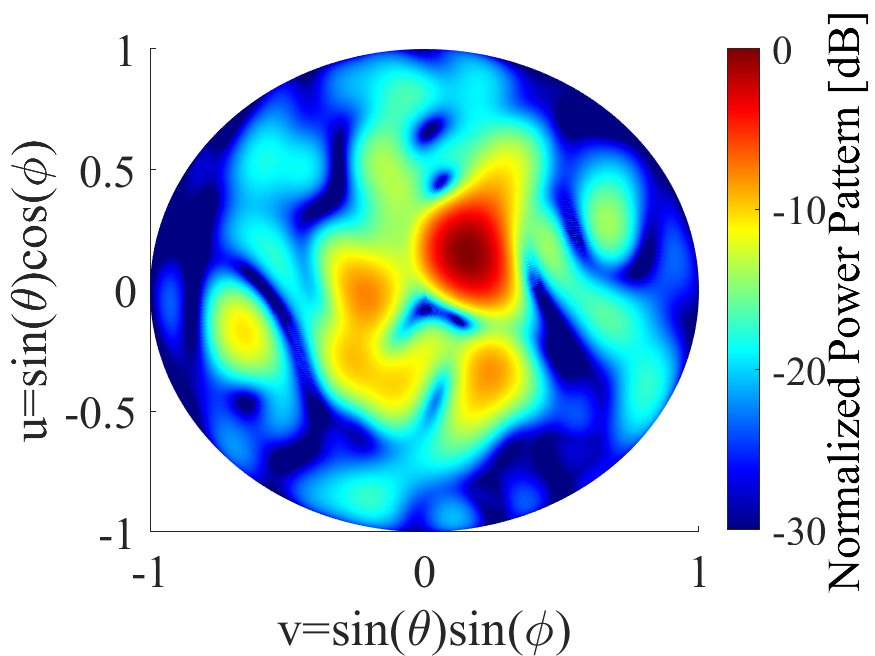}
    \includegraphics[width=0.32\linewidth]{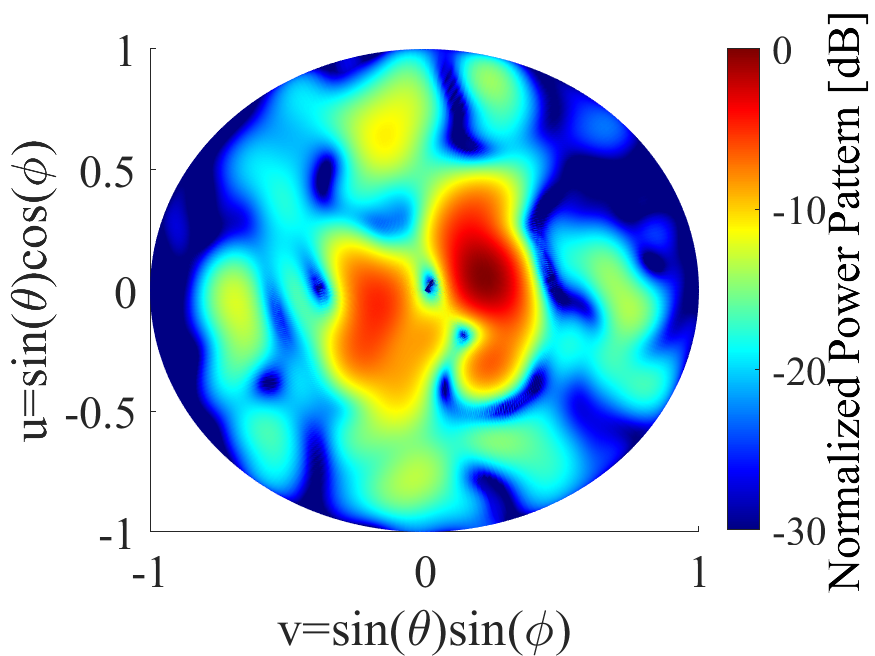}\\
    \small(g)\hspace{70pt}(h)\hspace{70pt}(i)\\
    \caption{Optimized $(N~\times~M)~=~8~\times~10$ TTPA, (a) Excitation amplitude, (d) excitation phase, and (g) normalized power pattern of FD architecture, (b) excitation amplitude, (e) excitation phase, (h) normalized power pattern of HFC architecture, (c) excitation amplitude, (f) excitation phase, and (i) normalized power pattern of HPC architecture, respectively.}
    \label{fig:TTPA}
\end{figure}

Fig.~\ref{fig:TTPA} depicts the detailed array excitation and normalized power pattern of optimized $(N~\times~M)~=~8~\times~10$ TTPA architectures. Figs.~\ref{fig:TTPA}(a),(b), and (c) depict the excitation amplitude, phase, and normalized power pattern for the TTPA with FD architecture, respectively. Figs.~\ref{fig:TTPA}(d),(e), and (f) depict the excitation amplitude, phase, and normalized power pattern for the TTPA with HFC architecture, respectively. Figs.~\ref{fig:TTPA}(g),(h), and (i) depict the excitation amplitude, phase, and normalized power pattern for the TTPA with HPC architecture, respectively.

We observe that thinned arrays maintain the two beams towards two receivers thanks to accurate phase control per radiating element. However, control on the SLLs is highly reduced due to more than half wavelength spacing. On the contrary, TTPAs maintain low SLLs but reduced control on the two beams towards two receivers due to amplitude and phase control on the subarray level. 

SLL and sum SE performance comparison of different irregular array architecture options in Figs.~\ref{fig:FPRAres},\ref{fig:ThinS40},~\ref{fig:ThinS20},  ~\ref{fig:DTPA}, and ~\ref{fig:TTPA} are listed in Table~\ref{Comptab}. In general, clustered arrays DTPA and TTPA achieve higher SE compared to their thinned array counterparts, owing to more radiating elements, respectively. FPRA with FD and HFC architectures exhibit high SE and low SLLs, but the hardware complexity is too high. FPRA with HPC architecture has poor SLLs due to reduced control per radiating patch, resulting in wider beams and higher SLLs. 50\% thinned arrays with FD and HFC architecture achieve similar SE with better SLLs compared to FPRA with HPC architecture. On the other hand, 25\% thinned arrays have poor SE due to fewer radiating elements and poor SLLs due to increased sparsity. DTPA with FD and HFC architectures maintains low SLLs and high SE thanks to the same number of radiating elements as FPRA becoming an excellent alternative to conventional FPRA architectures. TTPA with FD and HFC architectures maintains acceptable SLLs while significant reduction on the sum SE due to a combination of increased feeding losses $P_L$ and reduced array gain compared to DTPAs.
\begin{table}[]
\caption{SLL and sum SE performance comparison of different irregular array architectures in Figs.~\ref{fig:FPRAres},~\ref{fig:ThinS40},~\ref{fig:ThinS20},~\ref{fig:DTPA} and ~\ref{fig:TTPA} at $\eta~=~5$ dB serving two receivers ($K~=~2$).}\label{Comptab}
\centering
\begin{tabular}{|lcc|}
\hline
Architecture & SLL [dB] & sum SE [b/s/Hz]\\
\hline
FPRA FD    & -16.18  & 5.22     \\
FPRA HFC   &-16.52   &  5.22    \\
FPRA HPC   & -4.15  &  3.67    \\
\hline
50\% Thinned Array FD &  -9.57 &  3.37    \\
50\% Thinned Array HFC & -9.66 &  3.40    \\
50\% Thinned Array HPC & -4.53 &  2.14    \\
\hline
25\% Thinned Array FD & -4.98  & 2.09      \\
25\% Thinned Array HFC & -5.28 &   2.13   \\
25\% Thinned Array HPC &-3.06  & 1.24     \\
\hline
DTPA FD    &-12.89    & 4.53      \\
DTPA HFC   & -12.65   &  4.52    \\
DTPA HPC   & -5.64   &   3.04   \\
\hline
TTPA FD    & -8.98   &  3.67    \\
TTPA HFC   & -8.03   &  3.15    \\
TTPA HPC   &  -4.01  &  2.03    \\
\hline
\end{tabular}
\end{table}

\begin{figure}[]
    \centering
    \includegraphics[width=\linewidth]{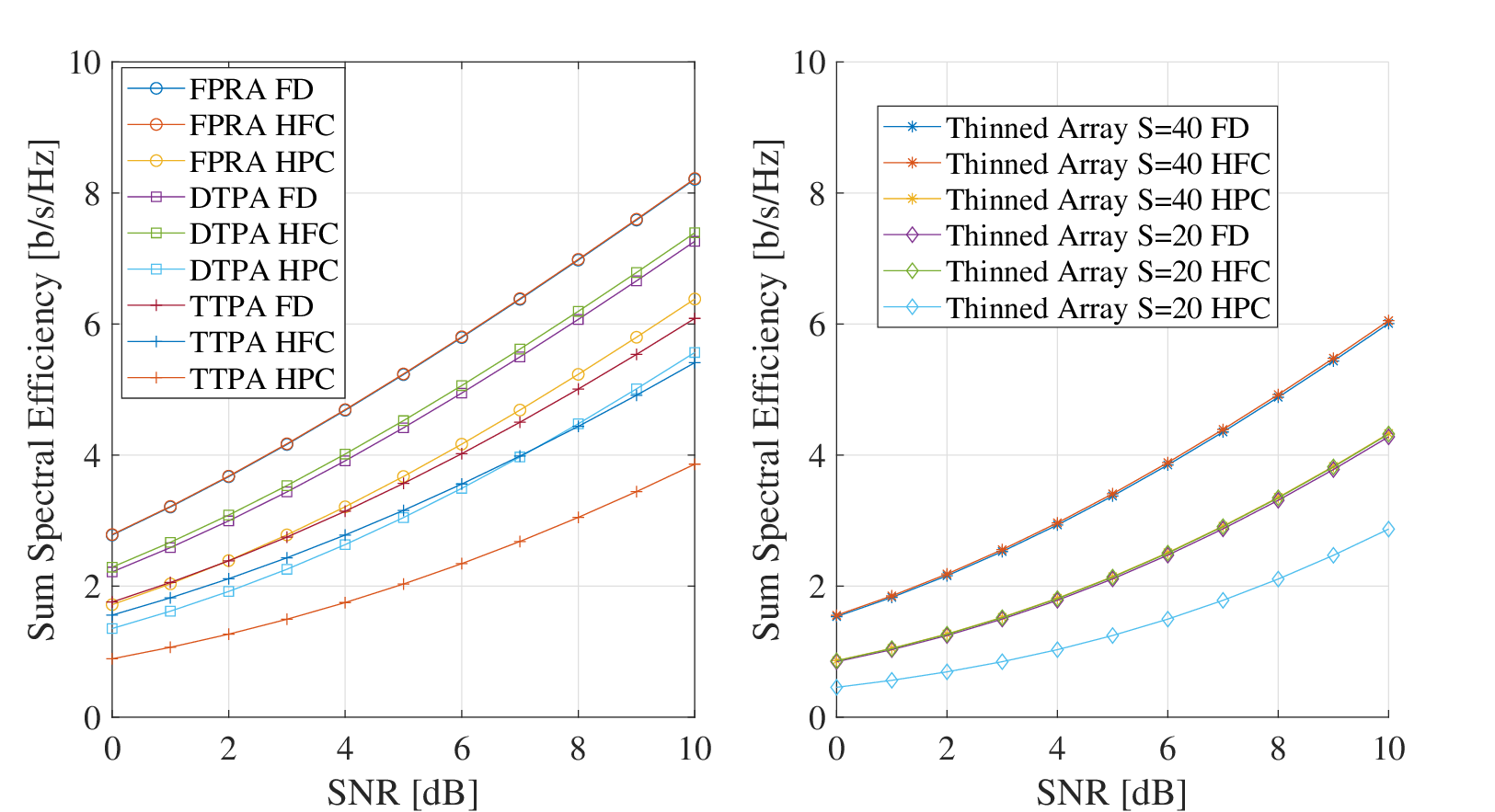}\\
    \small(a) \hspace{100pt} \small(b)\\
    \caption{SNR versus sum SE comparison of irregular arrays combined with different hybrid and digital architectures (a) FPRA, DTPA, and TTPA performances, (b) thinned array $S~=~40$, and $S~=~20$ performances, respectively.}
    \label{fig:snrvsSE}
\end{figure}

Finally, the sum SE versus SNR performance of different irregular array architectures depicted in Figs.~\ref{fig:FPRAres}, \ref{fig:ThinS40},\ref{fig:ThinS20}, \ref{fig:DTPA}, and \ref{fig:TTPA} have been compared considering 500 channel realizations.
Fig.~\ref{fig:snrvsSE} depicts the SNR versus sum SE comparison of irregular arrays combined with different hybrid and digital architectures. Similar to~\cite{Ertugrul2024_2}, clustered arrays outperform their thinned array counterparts, i.e., DTPA versus thinned array with $\rho~=~0.5$, and TTPA versus thinned array with $\rho~=~0.25$, respectively. We observe that the sum SE of FD and HFC architectures is almost the same for DTPA and thinned array configurations. DTPA with FD and HFC architecture outperforms the FPRA with HPC. On the other hand, TTPA with FD almost meets the FPRA HPC performance and outperforms DTPA with HPC. DTPA with FD architecture performance can be achieved with TTPA with FD architecture by increasing the PA output power $P_{PA}^{(out)}$ by 2~dB. An increase of 3~dB on $P_{PA}^{(out)}$ is required to meet the FPRA with FD performance for FPRA with HPC architecture. Similarly, DTPA with FD architecture performance can be met by DTPA with HPC architecture, having 3~dB increment on the $P_{PA}^{(out)}$.

\section{Conclusion}\label{sec:conc}
Beyond 100~GHz MIMO phased array systems require the utilization of irregular array architectures due to implementation challenges such as wavelength-IC size conflict. Hence, thinned and clustered array performances have been compared in the context of MU-MIMO targeting beyond 100~GHz communication systems. A new framework has been developed to compare the performance of irregular arrays in combination with hybrid/digital architectures. SLLs and sum SE were selected as the performance metrics to assess the communications performance and radiation characteristics simultaneously, while the antenna arrays have been excited with communication signals. An optimization problem was written to find the optimal irregular array architecture that exploits low SLLs and high sum SE simultaneously and allows arbitrary weighting of the two criteria. The number of configurations, i.e., arrangements and orientations of the antenna elements, grows significantly with the array aperture, which rules out brute force search methodologies. Hence, random sample generation schemes and GA based search methods have been utilized to generate many samples, and the optimal configuration has been selected among the generated samples. The optimal antenna array configurations have been realized in CST to validate the performance with full-wave simulations. The proposed methodology has been validated through numerical simulations where the performance of each array architecture is analyzed in detail, considering a MU-MIMO scenario with beyond 100~GHz channel models.

Simulation results showed that the clustered arrays DTPA and TTPA are outperforming their thinned array counterparts in terms of SLL and sum SE. In all cases, we observe significant performance reduction in HPC architecture compared to FD and HFC architectures due to limited amplitude and phase control per antenna. DTPA with FD or HFC architecture offers an excellent substitute for the conventional FPRA architectures. Increasing the clustering level from two to four by utilizing TTPA architectures resulted in acceptable SLLs but lower sum SE due to a combination of increased feeding losses and reduced array gain.

\bibliographystyle{IEEEtran}
\bibliography{mybib}

\end{document}